\documentclass[nologo,11pt,a4paper]{ETHpaper}

\usepackage{tabularx}
\usepackage{MnSymbol}
\usepackage{booktabs}
\usepackage{caption}
\usepackage{hhline}
\usepackage[table,dvipsnames]{xcolor}
\usepackage{graphicx}
\usepackage[english]{babel}
\usepackage[numbers,sort&compress]{natbib} 
\usepackage{setspace}

\title{Enhanced or distorted wisdom of crowds? \\ An agent-based model of \\ opinion formation under social influence}
\titlealternative{Enhanced or distorted wisdom of crowds?  \\ An agent-based model of opinion formation under social influence} 
\author{Pavlin Mavrodiev, Frank Schweitzer}
\authoralternative{P. Mavrodiev, F. Schweitzer}
\address{Chair of Systems Design, ETH Zurich, Switzerland\\
  \url{www.sg.ethz.ch}}
\www{\url{http://www.sg.ethz.ch}}

\reference{\emph{Submitted for publication}} 
\makeframing

\usepackage{appendix}
\usepackage{amsmath} \usepackage[utf8]{inputenc}
\usepackage{xcolor} \usepackage{soul}
\usepackage[hang]{subfigure}
\usepackage{multirow}
\usepackage{hyperref}
\usepackage{xparse,tikz,calc} \usepackage{pifont}

\renewcommand{\epsilon}{\varepsilon}

\newcommand{\mean}[1]{\left\langle #1 \right\rangle}
\newcommand{\abs}[1]{\left| #1 \right|}

\begin{document}

\maketitle

\begin{abstract}
  We propose an agent-based model of collective opinion formation to study the wisdom of crowds under social influence.
The opinion of an agent is a continuous positive value, denoting its subjective answer to a factual question. 
The wisdom of crowds states that the average of all opinions is close to the truth, i.e. the correct answer.
But if agents have the chance to adjust their opinion in response to the opinions of others, this effect can be destroyed.
Our model investigates this scenario by evaluating two competing effects: (i) agents tend to keep their own opinion (individual conviction $\beta$), (ii) they tend to adjust their opinion if they have information about 
  the opinions of others (social influence $\alpha$).
  For the latter, two different regimes (full information vs. aggregated information) are compared.
  Our simulations show that social influence only in rare cases enhances the wisdom of crowds.
  Most often, we find that agents converge to a collective opinion that is even farther away from the true answer.
  So, under social influence the wisdom of crowds can be systematically wrong.
 \end{abstract}

\section{Introduction}

Problems of collective decisions are tackled in different scientific disciplines, from biology to sociology, from computer science to management science, from robotics to statistical physics.
But the overarching research interest is the same: Collective decision processes should, in the best case, ensure an unanimous outcome, often denoted as \emph{consensus} \citep{Bose_2017}.
Taking a social perspective, individuals likely have deviating opinions regarding a particular issue.
But thanks to social interactions during the process of collective opinion formation, individuals are  able to adapt their opinions such that eventually a majority converges to the same opinion.
This ideal picture is distorted in reality in several ways.
To put forward the problem addressed in this paper: converging to the \emph{same} opinion does not imply converging to the \emph{right} opinion.
Collective decision processes can easily converge to a consensus that is wrong in an objective sense,
because of social influence among individuals. 

The term \emph{social influence} encompasses several aspects that are underestimated, or not yet understood regarding their consequences for opinion dynamics \citep{flache2017}.
For example, what is the role of social pressure, external influences, or affective involvement in collective decisions? 
Of particular importance is the question how individuals respond to the information they obtain about the opinions of others.
This question will be further addressed in our paper. 
For a systematic investigation, we utilize agent-based models that allow to vary such responses and to study their impact on the collective opinion.

Many agent-based models of opinion dynamics use a binary characterization of opinions, e.g. $\{0,1\}$ or $\{-1,+1\}$.
A particular class of stochastic models, the so-called ``voter models'' \citep{dornic2001critical,castellano2009statistical}, study how agents respond to the frequencies of these opinions in a predefined neighborhood \citep{Banisch2014,fernandez2014voter}.
In the linear voter model, the majority rule assumes that the opinion of the (local) majority is adopted with a probability directly proportional to the frequency of a given opinion. 
It can be shown analytically that such assumptions usually result in consensus.  
The question thus is how long it takes before this outcome is reached \citep{suchecki2005vmd}. 
Interestingly, a slow-down of the opinion dynamics on the agent level, i.e. a certain reluctance to change the opinion, can lead to a speed up of the opinion dynamics on the systemic level: consensus is reached faster  \citep{STARK2008}.
Also, the topology of the social network that facilitates agent interactions has an important influence on the consensus formation \citep{min2017fragmentation}, as well as possible nonlinear responses to the frequency of opinions (e.g. minority voting, voting against the trend, etc.) \citep{fs-voter-03}.
Eventually, such models can be extended to include assumptions of social impact theory \citep{nowak-szam-latane-90,lewenst-nowak-latane-92}, such as different weights of opinions, persuasion from agents with different opinions and support from agents with the same opinion \citep{holyst2001}.

The advantage of this model class is its analytical tractability, in addition to computer simulations, its disadvantage is in the many simplifications made \citep{schweitzer2018}.
More realistic models are based on a continuous characterization of opinions, e.g. as positive real numbers in a given interval, $[0,1]$.
The most prominent class of such models is the ``bounded confidence model'' that assumes agents only interact if the difference in their opinions is less than a certain threshold $\epsilon$ \citep{deffuant2001mixing,hegselmann2002opinion}. 
In this case, agents converge toward the mean of their opinions because of their interactions. 
The possibility to reach consensus then depends on the value of $\epsilon$ \citep{lorenz2007}. 
To foster consensus, assumptions about the topology of social interactions (e.g. group interactions) \citep{Meng_2018}, hierarchical decisions \citep{Pfitzner2013}, the influence of in-groups (agents known from previous interactions) \citep{Groeber2009} can be taken into account.

Still, this model class is rather abstract, focusing on effects on the conceptual level.
Towards more realistic models, the next step is to consider multi-dimensional opinions \citep{Schweighofer_2020,schweighofer20}.
Usually an individual's opinions on different subjects are not independent, but show correlations such that, out of a multi-dimensional opinion space, characteristic dimensions, e.g. \emph{left-right}, \emph{conservative-liberal} emerge \citep{schweighofer20}.
In addition to the simple attractive force assumed in the bounded confidence model (convergence toward the mean), now we have to consider also repulsive forces: agents increase the distance of their opinions as a result of their interaction.
In political space, this often leads to polarized opinions \citep{Banisch2019,mas2013differentiation}. 
Consensus in such scenarios is a rare exception, studies rather focus on the possibility to obtain qualified majorities 
\citep{baldassarri2007dynamics,garcia2012b,bornschier2015new}. 
Such models, while more complex, already allow to capture relations in real-world surveys.
They can be further enhanced by including the affective response of individuals and emotional influences \citep{Schweitzer_2020,schweighofer20}, in addition to social ones.

In conclusion, to formally model collective decision processes, we can build on a number of agent-based approaches at different levels of complexity.
In this paper, we will use the second model class described above, which uses a continuous opinion representation and a mutual convergence of individual opinions, to describe a particular phenomenon of collective decisions, namely the \emph{wisdom of crowds} (WoC). 
It applies to scenarios with factual questions where a \emph{true opinion} $\mathcal{T}$ exists, even if it is not known to individuals \citep{Galton1907}.
Our running example: ``How long is the border between Switzerland and Italy in kilometers?'' \citep{Lorenz2011a} has a true answer (734 km), while individuals usually only have raw estimates.
The WoC effect then states that the \emph{average} over all individual opinions is remarkably close to the true opinion.
So, it denotes a purely statistical effect but, interestingly, there is a lot of empirical evidence for it \citep{Kittur2008,Ray2006,Mannes2009}.
This is quantified by the \emph{small} collective error, $\mathcal{E}$, which is the squared difference between the average and the true opinion.

According to \citet{Surowiecki2005}, four criteria are required to form a \emph{wise crowd}: diversity of opinions, independence of opinions, some expertise, and the ability to aggregate individual opinions into collective opinions.
``Expertise''  only refers to the basic ability to give a meaningful answer to the question, e.g. the border of Switzerland is neither 0.1 kilometers, nor 3 million miles.
The ability to aggregate implies that someone, e.g. the social planner, indeed has access to all individual opinions, to calculate the aggregate.
The other two criteria are more important to us.
The diversity of opinions is \emph{large} only if individuals have \emph{independent} opinions.
But what happens if individuals obtain information about the opinions of others and then have the opportunity to revise their own opinion?
This question was investigated in a large empirical study \citep{Lorenz2011a}, where individuals had to answer the same question consecutively a number of times, while receiving different information from others.
In the so-called aggregated-information regime, they obtained at each time step information about the \emph{aggregated} opinion of others.
In the full-information regime, they instead obtained at each time step information about each \emph{individual} opinion.
These two information regimes allowed them to revise their own opinion accordingly. 

The study had two important findings:
(i) In the presence of information about the opinions of others, the group diversity of opinions drastically decreased.
That means, individuals tried to converge with their opinions, i.e. subjectively they got the impression to reach consensus.  
(ii) Despite this convergence, the collective error, which measures the distance to the truth, did \emph{not} decrease, but increased.
That means, the group collectively converged to an opinion which was objectively the wrong one.
This scenario, which was obtained quite frequently, is very dangerous because the group, converging to a common opinion, was collectively convinced that the wrong opinion was the right one.
This effect was stronger in the full-information regime compared to the aggregated-information regime. 

To answer the question, under what conditions this will happen, we need an agent-based model that reflects these two different information regimes and further allows to vary the strength of the social influence of other opinions, in comparison to the own conviction.
The aim of our paper is to provide such a model, as an extension of the bounded confidence model.
In particular, we want to model the full-information regime, which has shown the strongest effects, while the aggregated-information regime is used as a reference case. 

The rest of the paper is organised as follows.
Section \ref{model-description} introduces our agent-based model of opinion dynamics and relates it to macroscopic indicators to quantify the WoC effect.
Section \ref{computer-simulations} then presents the results of extensive agent-based simulations with respect to the three macroscopic measures of the wisdom of crowds: the collective error, the group diversity of opinions, and the WoC indicator.
Finally, the main conclusions are summarised in Section \ref{conclusion}.

\section{Modeling the wisdom of crowds}
\label{model-description}

\subsection{Micro dynamics of opinion formation}
\label{sec:micro-dynam-opin}

With the term ``micro dynamics'' we refer to the dynamics of the system elements, i.e. the agents.
We build on the framework of \emph{Brownian agents} \citep{agentbook-03}, which considers that agents have a
\emph{continuous internal degree of freedom}.  In our case, this is the opinion $x_{i}(t)$,
different for each agent $i=1,...,N$.  We assume that the values of $x_{i}$ can be mapped to
non-zero positive real numbers, i.e. $x_{i}(t)>0$, but are not bound to a defined interval.  This
reflects the experimental situations described before \citep{Lorenz2011a}.

These opinions can change over time because of influences from other agents or simply because agents
change their mind.  $x_{i}(0)$ denotes the initial value.  We propose the following general
dynamics \citep{schweitzer2020}:
\begin{align}
  \label{eq:1}
  \frac{d x_{i}(t)}{dt}= - \beta x_{i}(t) + \frac{1}{N} \sum_{j} \mathcal{F}_{ij}(t) + S_{i}(t)
\end{align}
This dynamics resembles the Langevin equation to describe Brownian motion, therefore we call these
Brownian agents.  The term $-\beta x_{i}$ is a \emph{damping term}, i.e. it describes, in the
absence of other influences, a relaxation process toward zero at a time scale $\beta$.  The term
$S_{i}(t)$ is an additive stochastic force that describes \emph{random influences} on the dynamics
of opinions.  We assume that these fluctuations are centered around the initial opinion $x_{i}(0)$,
to reflect the fact that individual convictions about certain subjects are likely to be long-lived.
Hence, $\mean{S_{i}(t)} \propto x_{i}(0)$ up to a constant.  We find it more convenient to rewrite
the stochastic part as:
\begin{align}
  \label{eq:2}
  S_{i}(t)=\beta x_{i}(0)+ A \xi_{i}(t) \;; \quad \mean{\xi_{i}(t)=0}\,; \quad \mean{\xi_{i}(t)\xi_{i}(t^{\prime})}=\delta(t-t^{\prime})
\end{align}
$\xi_{i}(t)$ is Gaussian white noise, i.e. it is not correlated in time and zero on average, $A$
denotes the strength of the stochastic force and is set equal to all agents.

The term $\mathcal{F}_{ij}(t)$ eventually describes how the change of opinion of agent $i$ is
influenced by the opinion of other agents $j$.
To better understand its impact, let us first assume $\mathcal{F}_{ij}(t)\equiv 0$, i.e agents do not have any information about the opinions of others. 
We call this the \emph{no-information regime}.
It implies that agents update their opinions stochastically \emph{without} considering
any other information.
With Eqs.~\eqref{eq:1}, \eqref{eq:2} the opinion dynamics for this case simply reads:
\begin{equation}
  \label{estimates0-no-info}
  \frac{dx_{i}(t)}{dt} = \beta\left[ x_{i}(0) - x_{i}(t) \right] + A \xi_{i}(t)  
\end{equation}
This stochastic equation denotes a standard Ornstein-Uhlenbeck process which also has an analytic solution \citep{pm-fs-analytic-20}. 
In this case the \emph{time average} of the individual opinion, $\overline{x_{i}(t)}$, equals $x_{i}(0)$ for large t with decreasing variance.

At difference with this trivial case, in this paper we discuss the case that agent $i$ has \emph{full
information} about the opinions of others and can take these into account in weighted manner, to
update her own opinion. This is reflected in the following assumption for $\mathcal{F}_{ij}(t)$:
\begin{align}
  \mathcal{F}_{ij}(t)= w_{ij}\, \left[x_{j}(t)-x_{i}(t)\right]
  \label{eq:4}
\end{align}
That means the social influence from the opinion of other agents increases with the difference
between opinions.  While this sounds like a simplified assumption, it has been empirically justified
in \citep{mavrodiev2013}, therefore we use it here.

The coupling variable $w_{ij}$ is chosen to be inversely proportional to the difference in
opinions:
\begin{align}
  w_{ij} = \dfrac{\mathcal{N}_{i}}{1+\exp
  \left\{\abs{x_{j}-x_{i}}/\alpha\right\} } \;;\quad
  \mathcal{N}^{-1}_{i}=\sum_{k=1}^{N}\dfrac{1}{1+\exp\left\{\abs{x_{k}-x_{i}}/\alpha\right\}}
  \label{eq:5}
\end{align}
with a normalization constant $\mathcal{N}_{i}$, such that $\sum_{j=1}^{N}w_{ij}=1$.
The parameter $\alpha$ acts as a measure of the strength of the social influence in the population.
Small values for $\alpha$ indicate that agents are less susceptible to others' opinions.  Conversely,
large values for $\alpha$ imply that even large differences in opinons between agents do not matter
much, therefore agents can resist stronger social influence.
In the limit case that $\abs{x_{j}-x_{i}}$ can be ignored, which is discussed further below, we see from Eqn.~\eqref{eq:5} that for $\alpha\to 0$ also $w_{ij}\to 0$, which makes sense.
Further, for the relevant range of values $0.2 \leq \alpha \leq 0.8$, $w_{ij}\propto \alpha$.

In summary, the stochastic dynamics for the opinion of agent $i$ in the full information regime
becomes:
\begin{align}
  \label{full-info}
  \frac{dx_{i}(t)}{dt} = \frac{1}{N}\sum_{j=1}^{N} \dfrac{\mathcal{N}_{i}}{1+\exp  \left\{\abs{x_{j}-x_{i}}/\alpha\right\}}
\left[x_{j}(t)-x_{i}(t)\right]
  + \beta 
  \left[x_{i}(0)-x_{i}(t)\right]+A\xi_{i}(t)
\end{align}
To make the impact of individual conviction, represented by the second term including $\beta$, comparable to the impact of social influence, represented by the first term including $\alpha$, we choose parameter values $0
\leq \alpha \leq 1$, and $1\leq \beta\leq 2$.
The above dynamics does not easily lend itself to analytical treatment, hence we will simulate it in
Section~\ref{computer-simulations}.  But to interpret the results, we first need to think about some comparison.

\subsection{Reference scenario}
\label{sec:reference-scenarios}

We want to compare our results to a reference case, in which agents have information only about the \emph{average opinion} of all other agents.
We call this the \emph{aggregated-information regime} \citep{Mavrodiev2012a,pm-fs-analytic-20}.
In this case the coupling variable $w_{ij}$ is effectively a
\emph{constant}, equal for all agents: $w_{ij}\equiv \alpha$.
Then, we can express the influence of other opinions as:
\begin{align}
  \frac{1}{N}  \sum_{j=1}^{N} \mathcal{F}_{ij}(t)=  \alpha \left[ \mean{x(t)}-x_{i}(t)\right]\;;\quad
  \mean{x(t)}=  \dfrac{1}{N}  \sum_{j=1}^{N} x_{j}(t)
  \label{eq:5a}
\end{align}
where $\mean{x(t)}$ is denoted as the \emph{mean} opinion in the following.  Eqn.~\eqref{eq:5a}
results in the stochastic dynamics:
\begin{equation}
  \label{estimates}
  \frac{dx_{i}(t)}{dt} = \alpha \left [\mean{x(t)} - x_{i}(t) \right] + \beta  \left[x_{i}(0) - x_{i}(t) \right] + A\xi_{i}(t)
\end{equation}
We note that Eq.~\eqref{estimates} bears similarities with the \emph{bounded confidence
  model} \citep{lorenz2007,hegselmann2002opinion,schweitzer2020}, in which agents update their opinion as follows:
\begin{align}
  \label{eq:3}
  \frac{dx_{i}(t)}{dt}= \frac{1}{N}\sum_{j} w^{\mathrm{b}}_{ij}\, \left[x_{j}(t)-x_{i}(t)\right]\;;\quad
  w^{\mathrm{b}}_{ij}=\alpha \,\Theta\left[z_{ij}(t)\right]\;;\quad
  z_{ij}(t)=\epsilon-\abs{x_{i}(t)-x_{j}(t)}
\end{align}
The coupling variable $w_{ij}^{b}$ reflects that agents only influence each other if the absolute
differences in their opinions is smaller than a confidence interval $\epsilon$.  $\Theta[x]$ denotes
the Heavyside function, i.e. $\Theta[x]=1$ only for $x\geq 0$ and $\Theta[x]=0$ otherwise.  This
means that agents consider the \emph{average opinions} of others, but take this average only over those
opinion not to far away from the own opinion. 

Compared to this case, our aggregated regime considers the information of all other agents on a
given agent and therefore is similar to a mean-field limit.  Additionally, the influence of the
initial opinion and stochastic effects are taken into account.

\subsection{Macroscopic measures for the wisdom of crowds}
\label{sec:macr-meas-wisd}

We remind that the wisdom of crowds (WoC) is expected to work if the diversity of individual opinions is
\emph{large}, while the deviation of the average opinion from the true value $\mathcal{T}$ is
\emph{small}.  Therefore, in line with previous studies \citep{Lorenz2011a}, we will use the \textit{collective error}, $\mathcal{E}$, the \textit{group diversity}, $\mathcal{D}$, and the \textit{wisdom of crowds indicator}, $\mathcal{W}$, as
macroscopic measures to evaluate these conditions.

In order to define these measures, we need to keep in mind that the distribution of opinions in the considered experimental setup is very broad, and can be proxied by a \emph{log-normal} distribution.
Therefore, the \emph{average opinion} is not 
well represented by the \emph{arithmetic} mean, $\mean{x}=\sum_{i}x_{i}/N$, but by the geometric mean, which is equivalent to the arithmetic mean of the $\log$ values, $\mean{\ln x}$, \citep{Lorenz2011a}.
Further, because opinions change over time according to Eqn.~\eqref{full-info}, we are only interested in the long-term (LT) values of these macroscopic measures.

The long-term collective error $\mathcal{E}_{\mathrm{LT}}$ shall be defined as the squared deviation of the average opinion from the true value, $\mathcal{T}$:
\begin{equation}
  \label{CE}
  \mathcal{E}_{\mathrm{LT}} = \left[\ln{\mathcal{T}} - \mean{\ln{x_{\mathrm{LT}}}} \right]^{2},
\end{equation}
The long-term group diversity $\mathcal{D}_{\mathrm{LT}}$ is given by the \emph{variance} of the distribution of opinions:
\begin{align}
  \label{Diversity}
  \mathcal{D}_{\mathrm{LT}}&= \mathrm{Var}\left[\ln x_{\mathrm{LT}}\right]= \dfrac{1}{N}\displaystyle\sum\limits_{i=1}^{N}\left[ \ln
                   x_{i_{\mathrm{LT}}}- \mean{\ln x_{\mathrm{LT}}}\right]^{2} = \mean{\left[\ln x_{\mathrm{LT}}\right]^{2}} - \mean{\ln
                   x_{\mathrm{LT}}}^{2}
\end{align}
The WoC indicator $\mathcal{W}$ is measured by the deviation of the truth from the long-term \emph{median opinion}, $\hat{x}_{\mathrm{LT}}$.
This means that for $N/2$ of all opinions $x_{i_{\mathrm{LT}}}\leq \hat{x}_{\mathrm{LT}}$ holds, and $x_{i_{\emph{LT}}}\geq \hat{x}_{\mathrm{LT}}$ for the remaining $N/2$.
In plain words, the value of $\mathcal{W}$  indicates how \emph{central} the position of true value $\mathcal{T}$ is within the distribution of opinions.
$\mathcal{W}$ achieves a maximum value if the truth can be bracketed by the two most central opinions.

More formally, if $X_{N}$ is the set of all $N$ opinions, we have the set of \emph{ordered} opinions, $\{\bar{x}_{i} | \bar{x}_{i} \in
X_{N}, \bar{x}_{i} \leq \bar{x}_{j}, ~\forall~i<j\}$.
The indicator $\mathcal{W}$ is defined as $\max\{i | \bar{x}_{i} \leq \mathcal{T} \leq \bar{x}_{N-i+1}\}$.
It reaches a maximum of $[N/2]$ when the truth is either the most central opinion or is bracketed by the two
most central opinions, and a minimum of 0 when $\mathcal{T} \notin (\bar{x}_{0},\bar{x}_{N})$.
To illustrate this, let us assume we have a set of 100 \emph{ordered} opinions such that $\{1,2,3,...,99,100\}$.
The two most central opinions in this set are 49 and 50, or 50 and 51, respectively.
If the true value would be 50, it is just bracketed by the two most central opinions.
This would give us $\mathcal{W}=50=N/2$.
If the true value would be 70, it is no longer bracketed by the two most central opinions.
Instead it is 20 positions away from the central opinions, hence $\mathcal{W}=30$.

\subsection{Initial configuration and numerics}
\label{sec:init-conf}

\begin{figure}[htbp]
  \centering
  \includegraphics[scale=0.2]{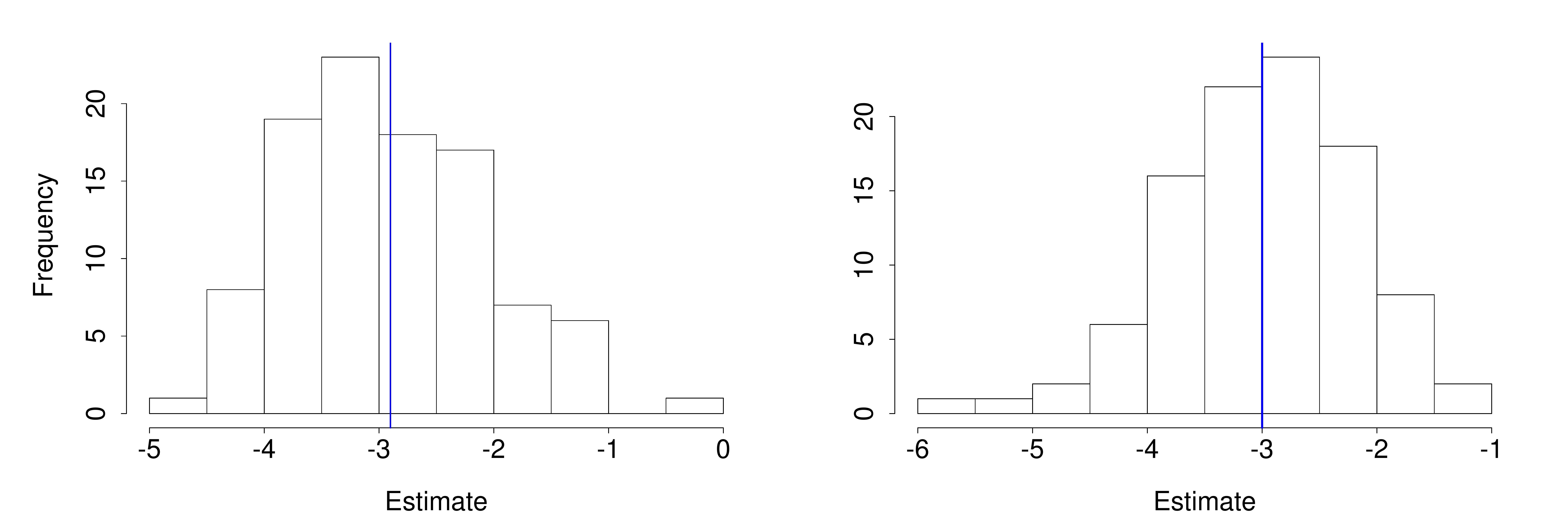}
  \caption{Histograms of initial opinions, $x_{i}(0)$, sampled from two different log-normal distributions. The blue lines indicate the different mean values
    $\mu^{(1)}_{\ln x}$=-2.9 (left) and $\mu^{(2)}_{\ln x}$=-3.0 (right), the variance $\sigma^{2}_{\ln x}$=0.72 is the same.
Note the \emph{logarithmic} values of the $x$-axis.}
  \label{init-cond}
\end{figure}

In order to perform agent-based simulations using Eqs.~\eqref{full-info}, \eqref{estimates} we need to specify the initial distribution of opinions, from which $N$ values $x_{i}(0)$ are sampled.
We take as an input two different log-normal distributions $P^{(1)}(x)$, $P^{(2)}(x)$ with two different mean values $\mu^{(1)}_{\ln x}$=-2.9 and $\mu^{(2)}_{\ln x}$=-3.0, but the same variance $\sigma^{2}_{\ln x}$=0.72.
The histograms of the $N=100$ opinions sampled from these two initial distributions are plotted in Figure~\ref{init-cond}. 

Further, we have considered three different true values, $\ln \mathcal{T}$=\mbox{-2.00}, $\ln \mathcal{T}$=-2.90, $\ln \mathcal{T}$=-3.12.
With these values and the parameters of the initial  distributions, we can determine the \emph{initial} collective errors $\mathcal{E}(0)$ from Eqn.~\eqref{CE} and  the \emph{initial} group diversity $\mathcal{D}(0)$ from Eqn.~\eqref{Diversity}.
With these specifications, we run $N=100$ agent-based simulations in parallel.
The dynamics are solved using the $4^{\text{th}}$ order Runge-Kutta method for the full-information regime, Eqn.~\eqref{full-info}, and the Euler method for the aggregated-information regime, Eqn.~\eqref{estimates}.
We used a constant time step $\Delta t=0.01$, and a final time $t=3000$, to obtain the long-term values.
The noise intensity was chosen as a rather small value, $A=10^{-3}$.

\section{Results of agent-based simulations}
\label{computer-simulations}

\subsection{Collective error}
\label{sec:reference-scenarios-1}

We present all our results as heat maps of the relevant quantities, $\mathcal{E}_{\mathrm{LT}}$, $\mathcal{D}_{\mathrm{LT}}$ or $\mathcal{W}_{\mathrm{LT}}$, dependent on the two model parameters, social influence, $\alpha$, and individual conviction, $\beta$, for which we have performed a thorough parameter sweep.
The \emph{right} column always refers to the full-information regime, the focus of our paper.
The \emph{left} column shows the matching results for the reference case, the aggregated-information regime, which has been  discussed in \citep{Mavrodiev2012a} regarding agent-based simulations and in \citep{pm-fs-analytic-20} regarding analytic solutions for the macroscopic measures.

\begin{figure}[htbp]
  \centering
  \includegraphics[width=0.45\textwidth]{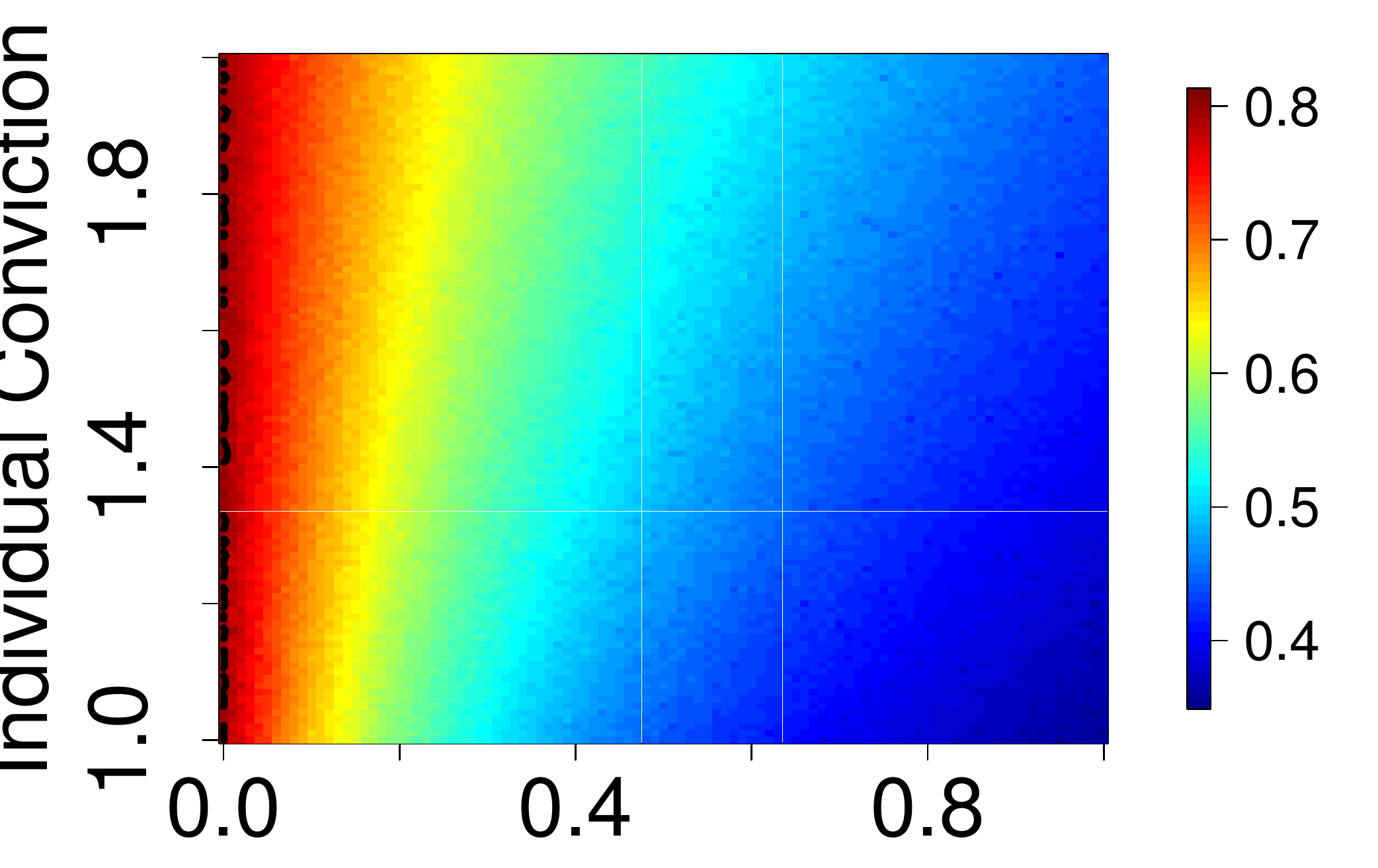}
  \hfill(a)\hfill
\includegraphics[width=0.45\textwidth]{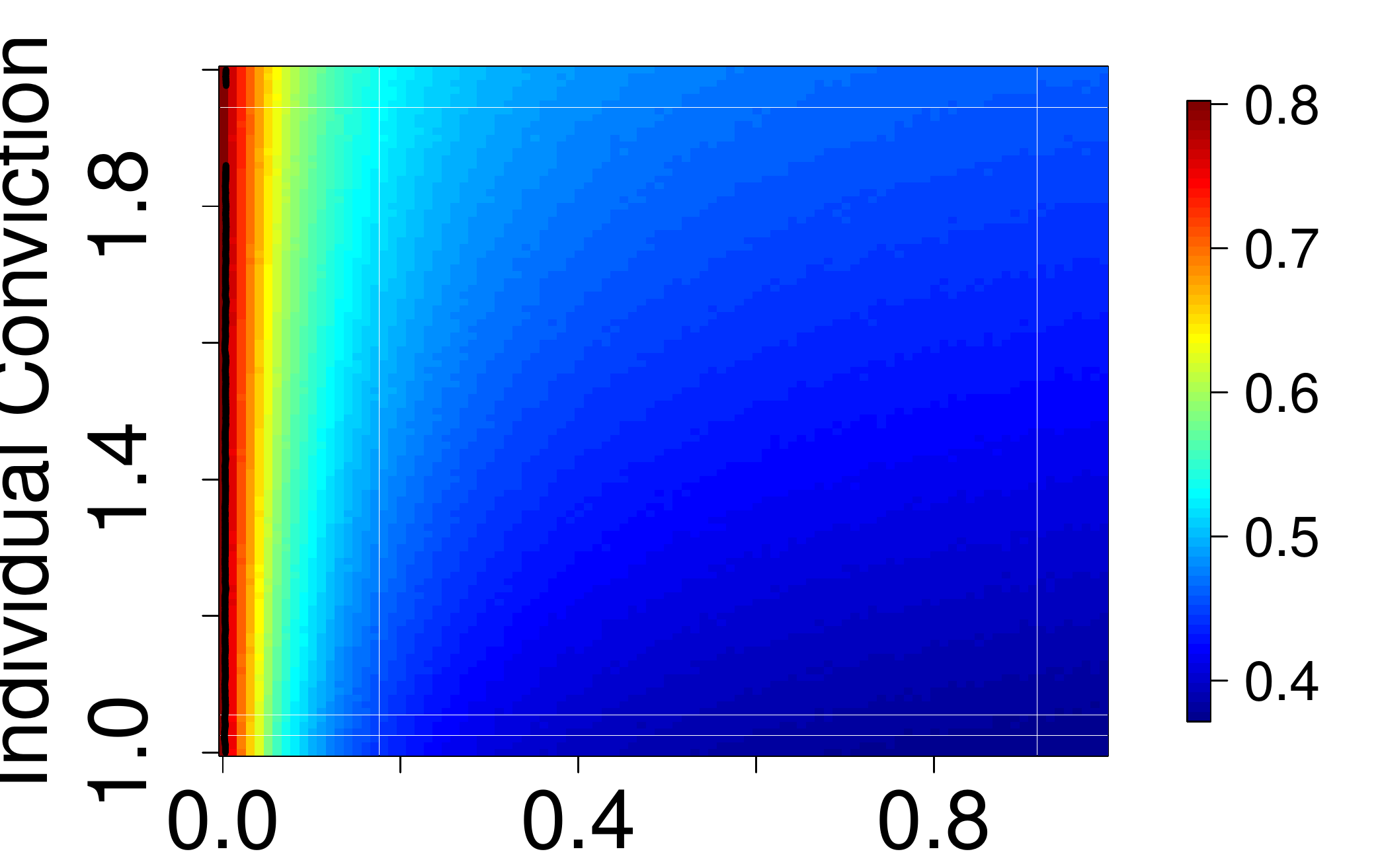} \\  
  \includegraphics[width=0.45\textwidth]{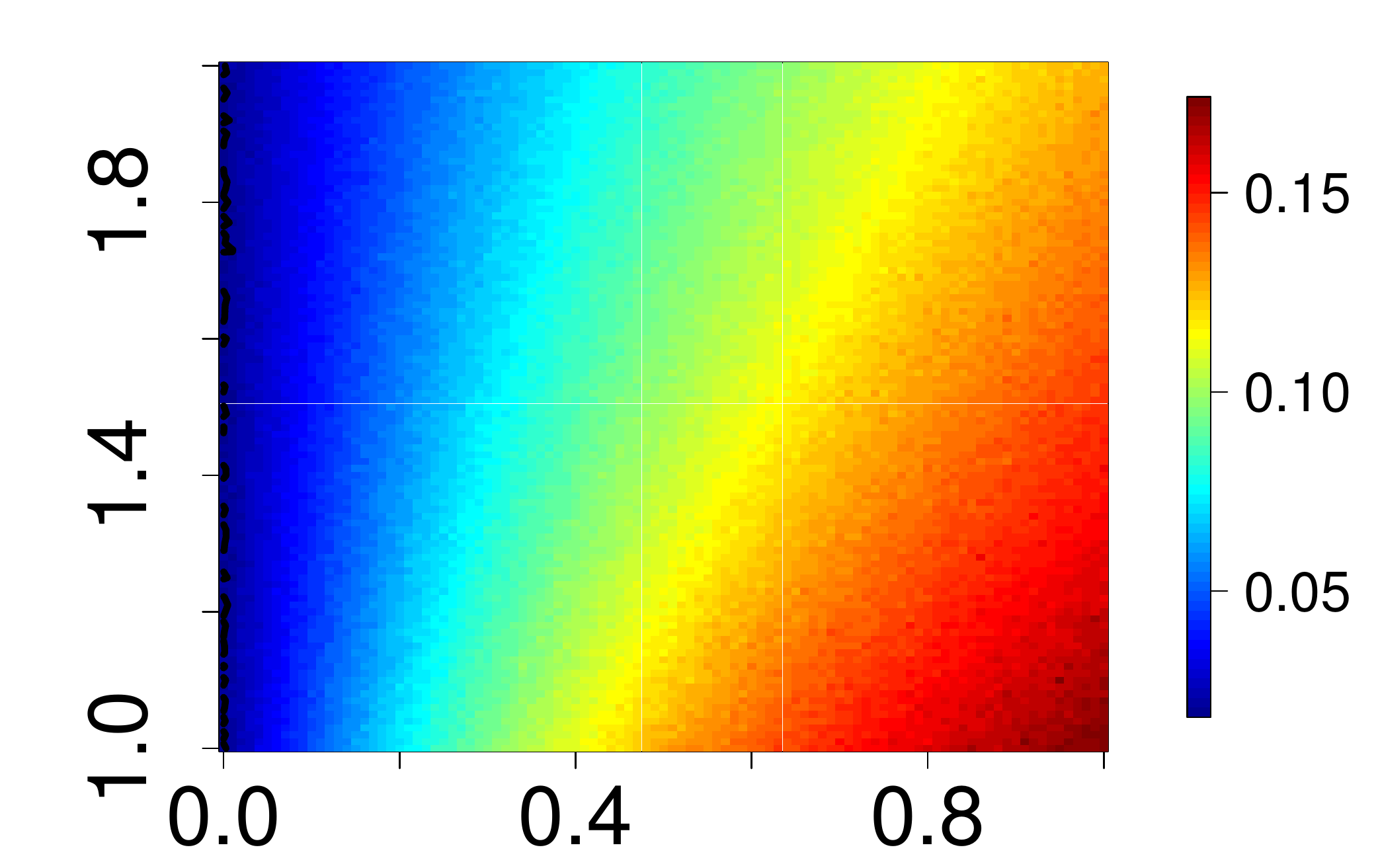}
\hfill(b)\hfill
\includegraphics[width=0.45\textwidth]{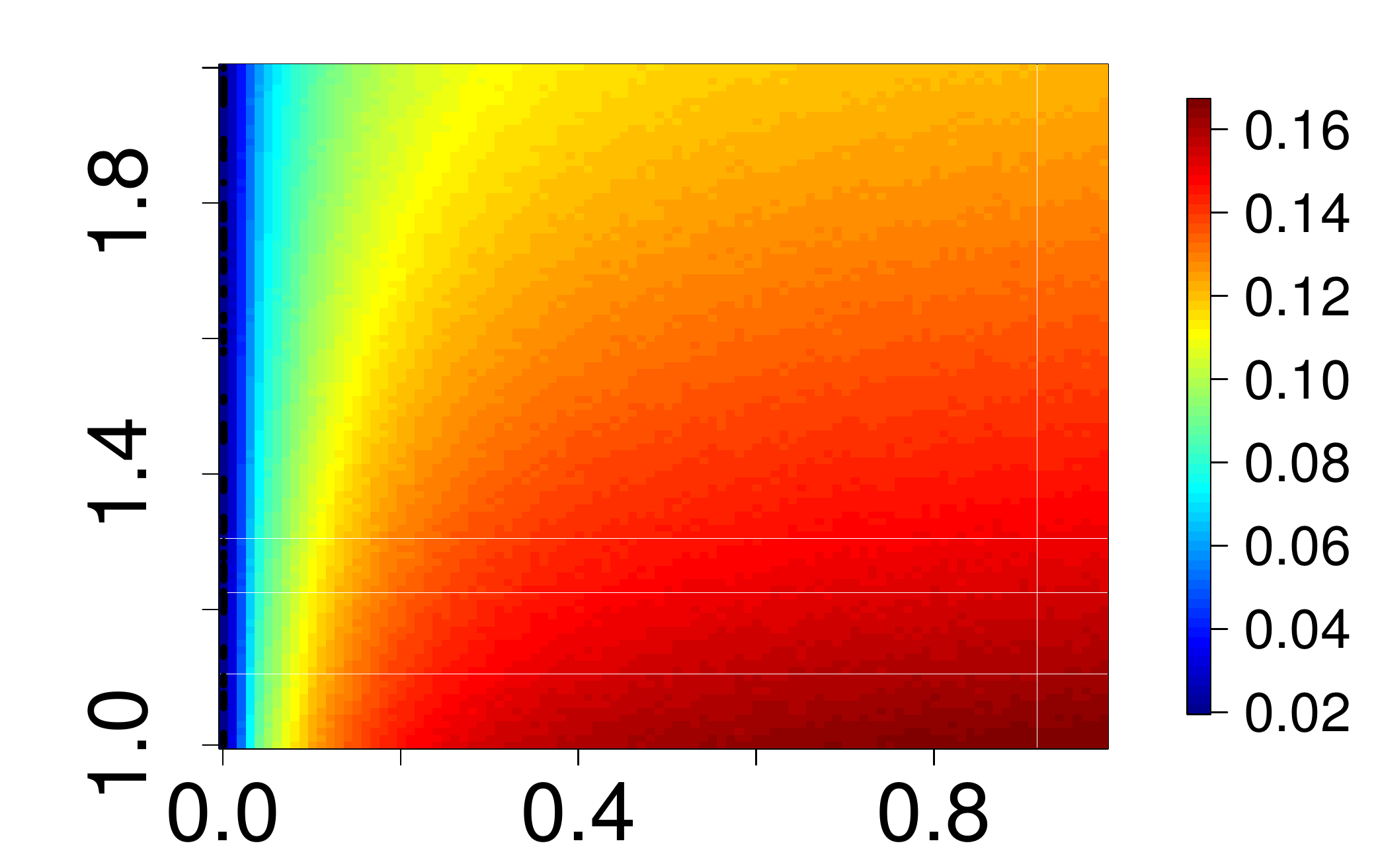}\\  
\includegraphics[width=0.45\textwidth]{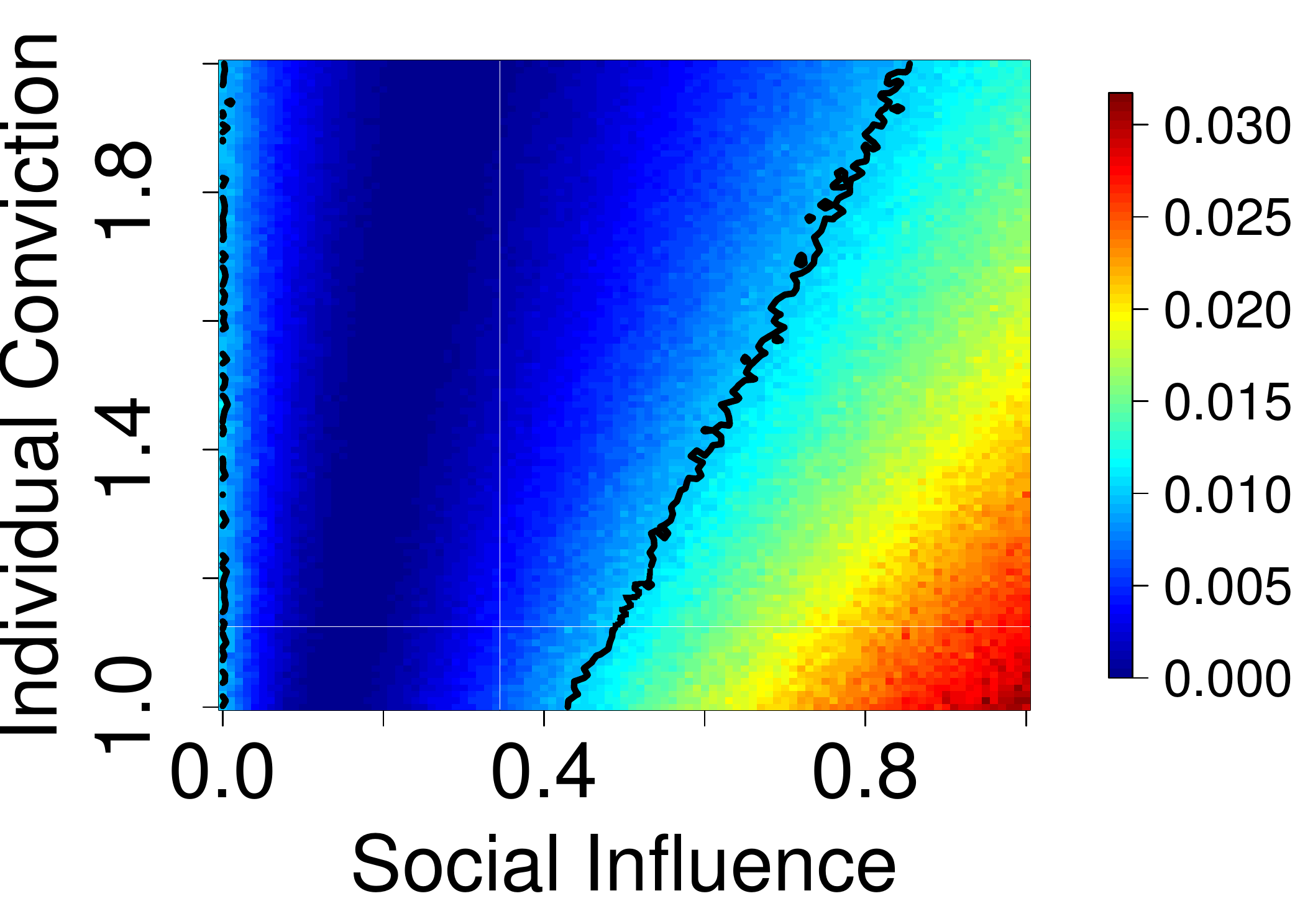}
\hfill(c)\hfill
\includegraphics[width=0.45\textwidth]{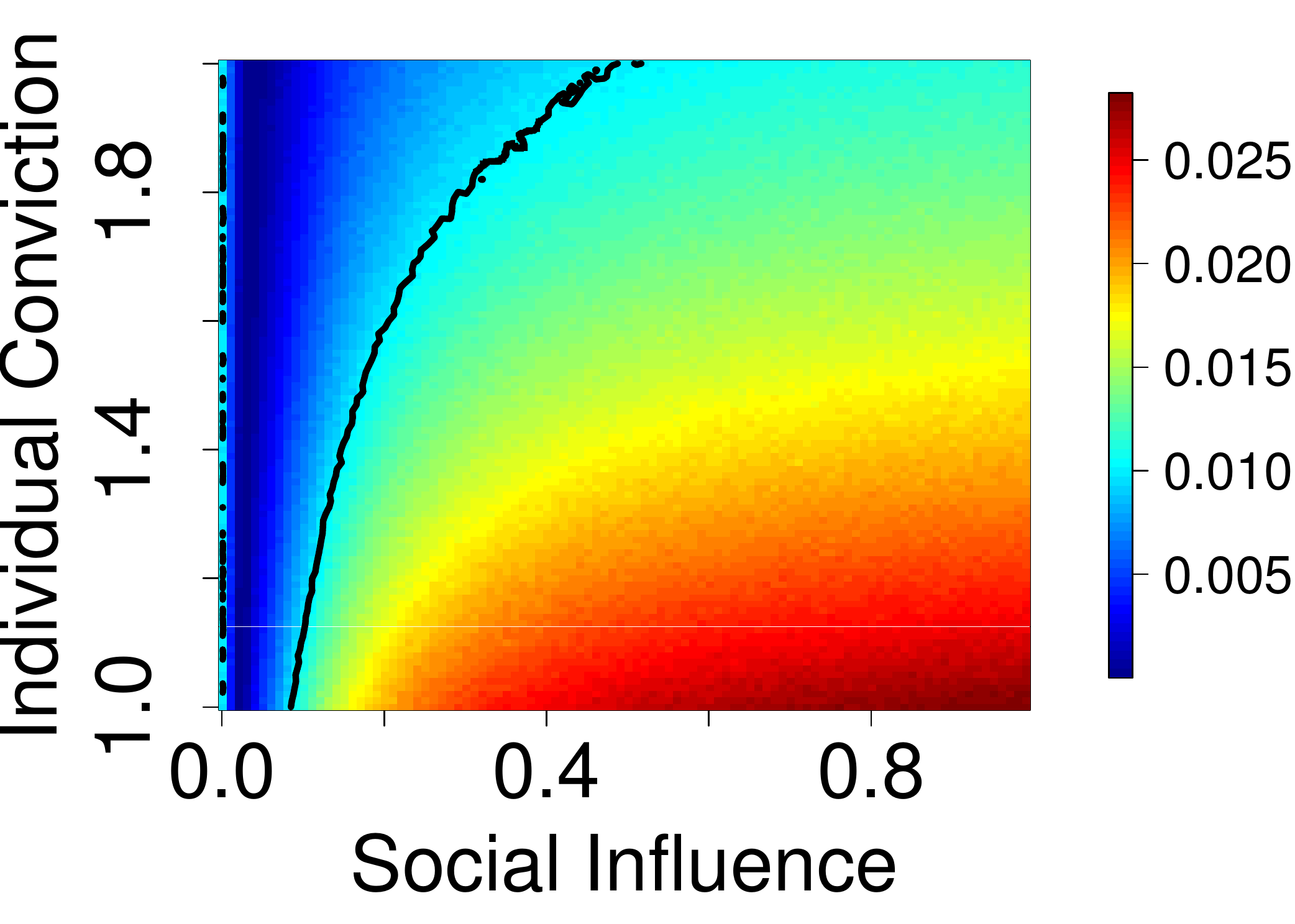}  
\caption{Agent-based simulations of the long-term collective error $\mathcal{E}_{\mathrm{LT}}$ (color coded) dependent on the values of social influence $\alpha$ ($x$-axis) and individual conviction $\beta$ ($y$-axis).
  \textbf{(left)} Aggregated-information regime, Eqn.~\eqref{estimates}, \textbf{(right)} full-information regime, Eqn.~\eqref{full-info}.
  Parameters: $N$=100, $t$=3000, $A$=$10^{-3}$, $\Delta t$=0.01. 
  Different initial conditions: 
  \textbf{(a)} $\mathcal{E}$(0)=0.80, $\ln \mathcal{T}$={-2.00}, $\mean{\ln x(0)}$=-2.9, 
  \textbf{(b)} $\mathcal{E}$(0)=0.02, $\ln \mathcal{T}$=-3.12, $\mean{\ln x(0)}$=-3.0, 
  \textbf{(c)} $\mathcal{E}$(0)=0.01, $\ln \mathcal{T}$=-2.90, $\mean{\ln x(0)}$=-3.0. 
  Black contour lines indicate regions in the parameter space where $\mathcal{E}_{\mathrm{LT}}=\mathcal{E}(0)$.
  Note that these are vertical lines at $\alpha$=0 in all plots, in (c) there is an additional line.
  }
  \label{sweep2}
\end{figure}

Figure~\ref{sweep2} shows the results for $\mathcal{E}_{\mathrm{LT}}$ for three different initial conditions (a)-(c).
The color code indicates the value of $\mathcal{E}_{\mathrm{LT}}$: (red) for high values, which is bad, and (blue) for low values, which is good.
We note that for each plot the color code represents \emph{different} values.
A  black line indicates parameter combinations $(\alpha,\beta)$ for which $\mathcal{E}_{\mathrm{LT}}=\mathcal{E}(0)$.
Mostly, these lines are not noticeable because they coincide with $\alpha=0$.

In Figure~\ref{sweep2}(a) the initial condition is chosen such that the average initial opinion is \emph{far away} from the true value, i.e. the initial collective error is \emph{high}.
Then, we find that an increase of social influence $\alpha$ considerably decreases the collective error, which means it improves the wisdom of crowds, whereas individual conviction has little impact on the outcome.
This positive finding is much stronger in the full-information regime, in which agents have access to the opinions of all other agents.
It decreases if they have access only to  the aggregated opinion.

The situation inverts if instead the initial condition is chosen such that the average initial opinion is \emph{close} to the true value, i.e. the initial collective error is \emph{low}.
Then an increase in social influence $\alpha$ \emph{can} worsen the outcome, leading the average opinion farther away from the true value, as shown in Figure~\ref{sweep2}(a). 
This is the most dangerous case: agents collectively converge to a common opinion, but this, from an objective perspective, is the wrong one.
This effect, again, is much stronger in the full-information regime.
In the aggregated-information regime, we could still identify parameter ranges with low $\alpha$, where the wisdom of crowds is not much distorted.
But more information for the agents destroys this possibility.
Individual conviction, as the second influential parameter, impacts the results mainly for the aggregated-information regime, but is less noticeable in the full-information regime.

This leads us to the question why there is this \emph{monotonous deterioration} of the wisdom of crowds with increasing social influence.
A theoretical investigation of the average opinion $\mean{\ln x(t)}$ in the aggregated-information regime  \citep{pm-fs-analytic-20} tells us that it can only increase over time, $d \mean{\ln x(t)}/dt >0$.
Therefore, if initially $\mean{\ln x(0)}< \ln \mathcal{T}$, there is a \emph{chance} that  $\mean{\ln x(t)} \to \ln \mathcal{T}$ over time, if $\alpha$ is large enough and $\beta$ is not too strong.
However,  if initially $\mean{\ln x(0)}> \ln \mathcal{T}$, the interaction dynamics of the agents can only lead their average opinion \emph{further away} from the true value.
This is the case shown in Figure~\ref{sweep2}(b), and it becomes worse both if social influence increases or if full information about the opinion of others is provided. 

Consequently, we should also expect situations \emph{without} a monotonous deterioration.
This is shown in Figure~\ref{sweep2}(c), which illustrates a non-monotonous dependence of the collective error on the social influence.
Here, the initial collective error is also low, as in (b), but now the initial distribution of opinions is such that $\mean{\ln x(0)}< \ln \mathcal{T}$.
Hence, we find that for low values of the social influence the collective opinion indeed converges to the true value (indicated by deep blue).
But the parameter range is rather small, and much smaller than in the aggregated-information regime.
Hence, we can conclude that only low social influence can improve the wisdom of crowds, whereas large social influence most likely deteriorates it.

The impact of the second model parameter, individual conviction $\beta$, becomes more visible for the chosen initial conditions (c).
Taking a fixed social influence, e.g. $\alpha$=0.5, we find that individual conviction can \emph{improve} the collective error, because it acts as a kind of reluctance \emph{against} converging to the (wrong) collective opinion. 

We also point to the visible black line which indicates the conditions under which the collective error does \emph{not} change, i.e. is equal to the initial value.
In the full-information regime, this is not a straight line as found for the reference case.
That means, non-linear effects become much stronger if agents have access to all other opinions.

\subsection{Group diversity}
\label{sec:group-diversity}

\begin{figure}[htbp]
  \centering
  \includegraphics[width=0.45\textwidth]{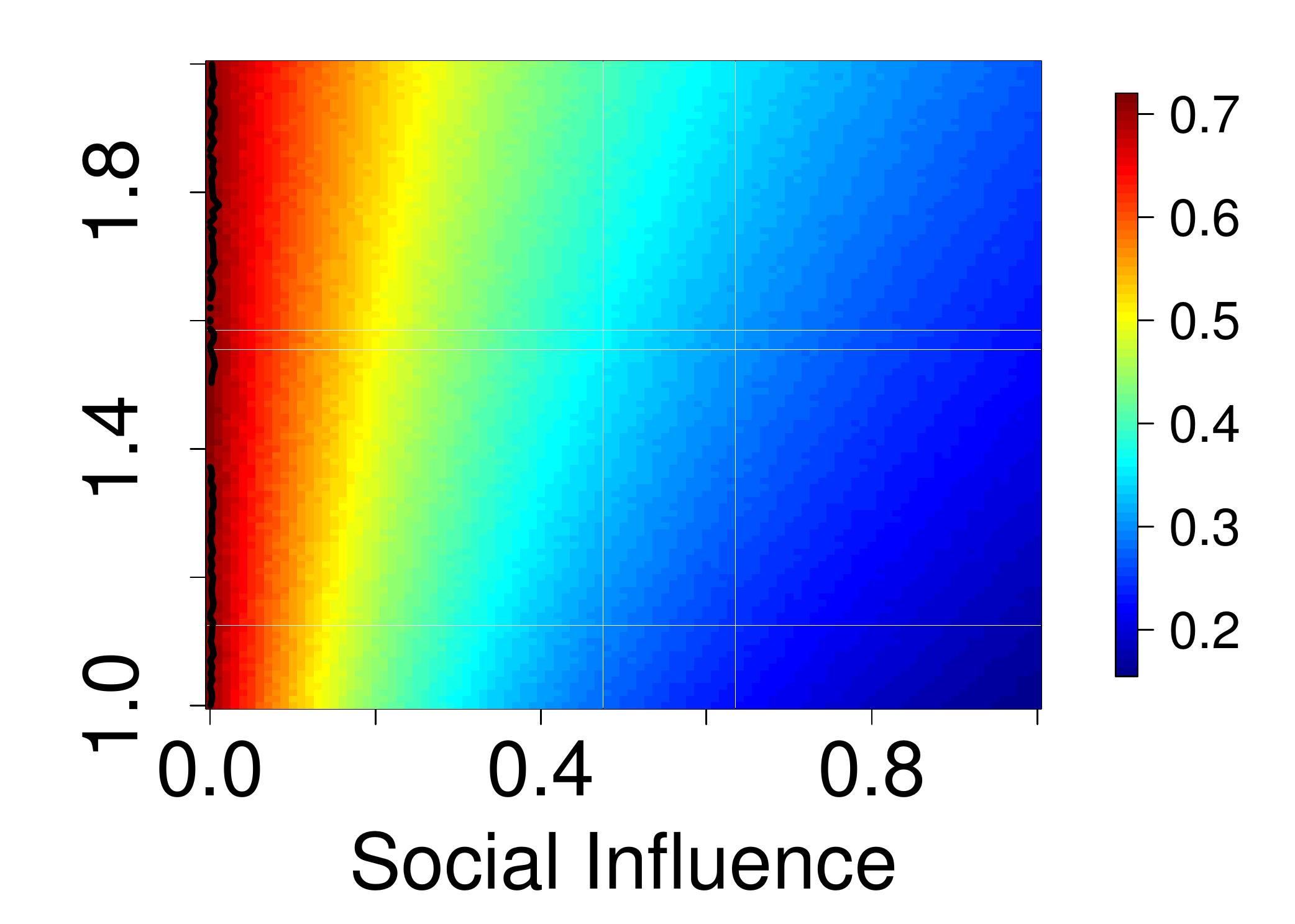}
  \hfill
\includegraphics[width=0.45\textwidth]{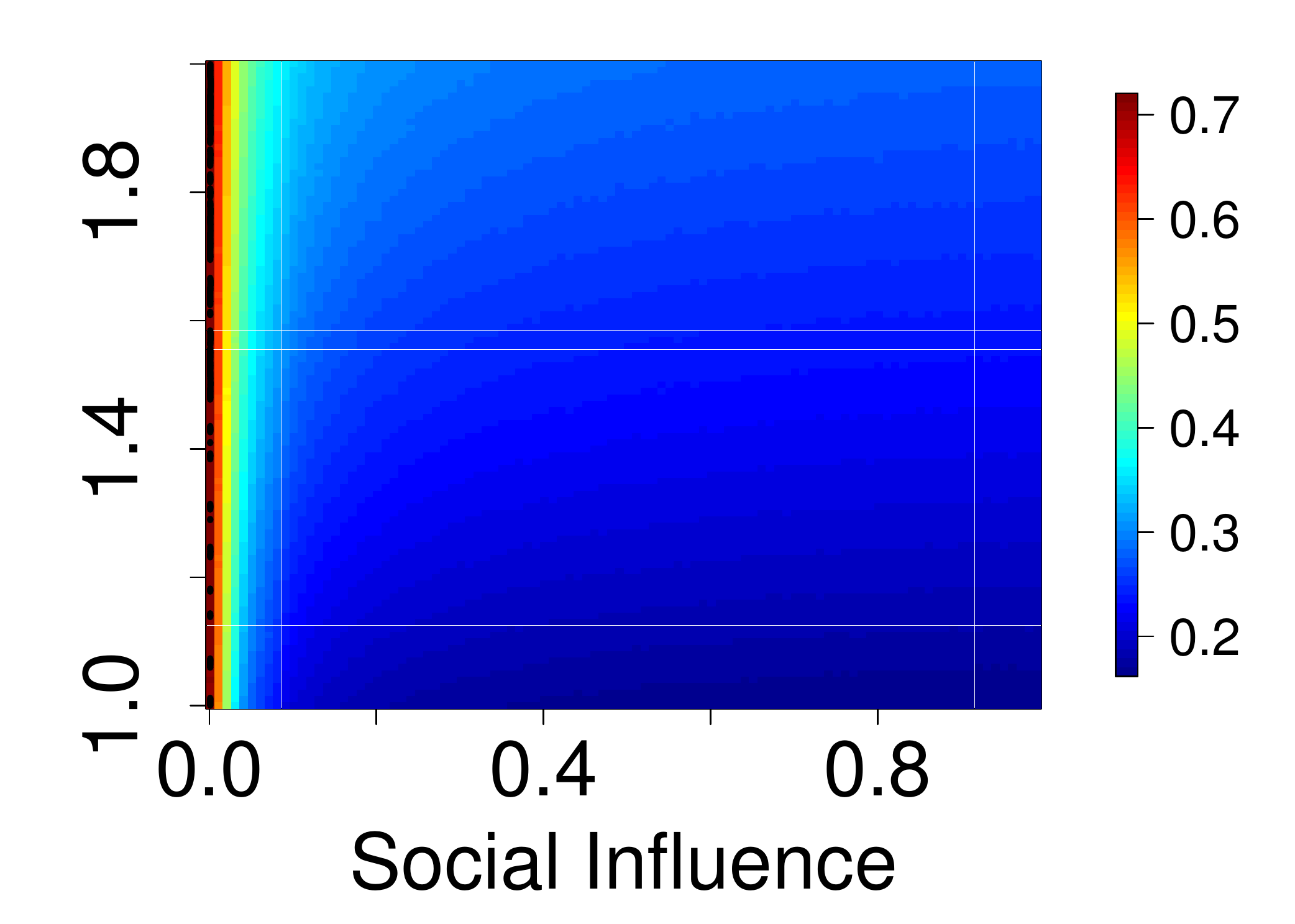}
  \caption{Agent-based simulations of the long-term group diversity $\mathcal{D}_{\mathrm{LT}}$ (color coded) dependent on the values of social influence $\alpha$ ($x$-axis) and individual conviction $\beta$ ($y$-axis).
  \textbf{(left)} Aggregated-information regime, Eqn.~\eqref{estimates}, \textbf{(right)} full-information regime, Eqn.~\eqref{full-info}.
  Parameters: $N$=100, $t$=3000, $A$=$10^{-3}$, $\Delta t$=0.01.  
  Initial condition: $\mathcal{D}$(0)=0.72.}
  \label{fig:groupdiv}
\end{figure}

The second macroscopic measure for the wisdom of crowds is shown in Figure~\ref{fig:groupdiv}, again in comparison to the reference case.
Note that the long-term group diversity $\mathcal{D}_{\mathrm{LT}}$ does \emph{not} depend on the initial conditions, but only on $(\alpha,\beta)$ and $\mean{\delta^{2}(0)}$.
Therefore the plot is the same for all initial conditions shown in Figure~\ref{sweep2}. 
The color scale is chosen such that a \emph{large} group diversity, which is good, is indicated by \emph{red}, whereas a \emph{small} group diversity is shown in \emph{blue}. 
We find that in the full-information regime the group diversity is drastically reduced by means of social influence.
Its high initial value cannot be maintained, not even for the case of a high initial conviction $\beta$. 

This becomes a problem for those parameter ranges, where a low group diversity is combined with a high collective error.
In this case, the agent population collectively converges to the (objectively) \emph{wrong} opinion.
We see that this is most likely  the case for the full-information regime, whereas in the reference case the parameter range for acceptable values of the group diversity is much larger.

\subsection{WoC indicator}
\label{sec:woc-indicator}

Finally we discuss the results for the long-term WoC indicator $\mathcal{W}_{LT}$, which measures
the distance between the truth and the median of the opinion distribution.
Ideally the true value should be ``central'' with respect to the opinion distribution, which means for the given configurations, it should have values around $[N/2]$=50.
The color code is chosen such that (red) indicates good values, (blue) a bad outcome.

\begin{figure}[htbp]
 \includegraphics[width=0.45\textwidth]{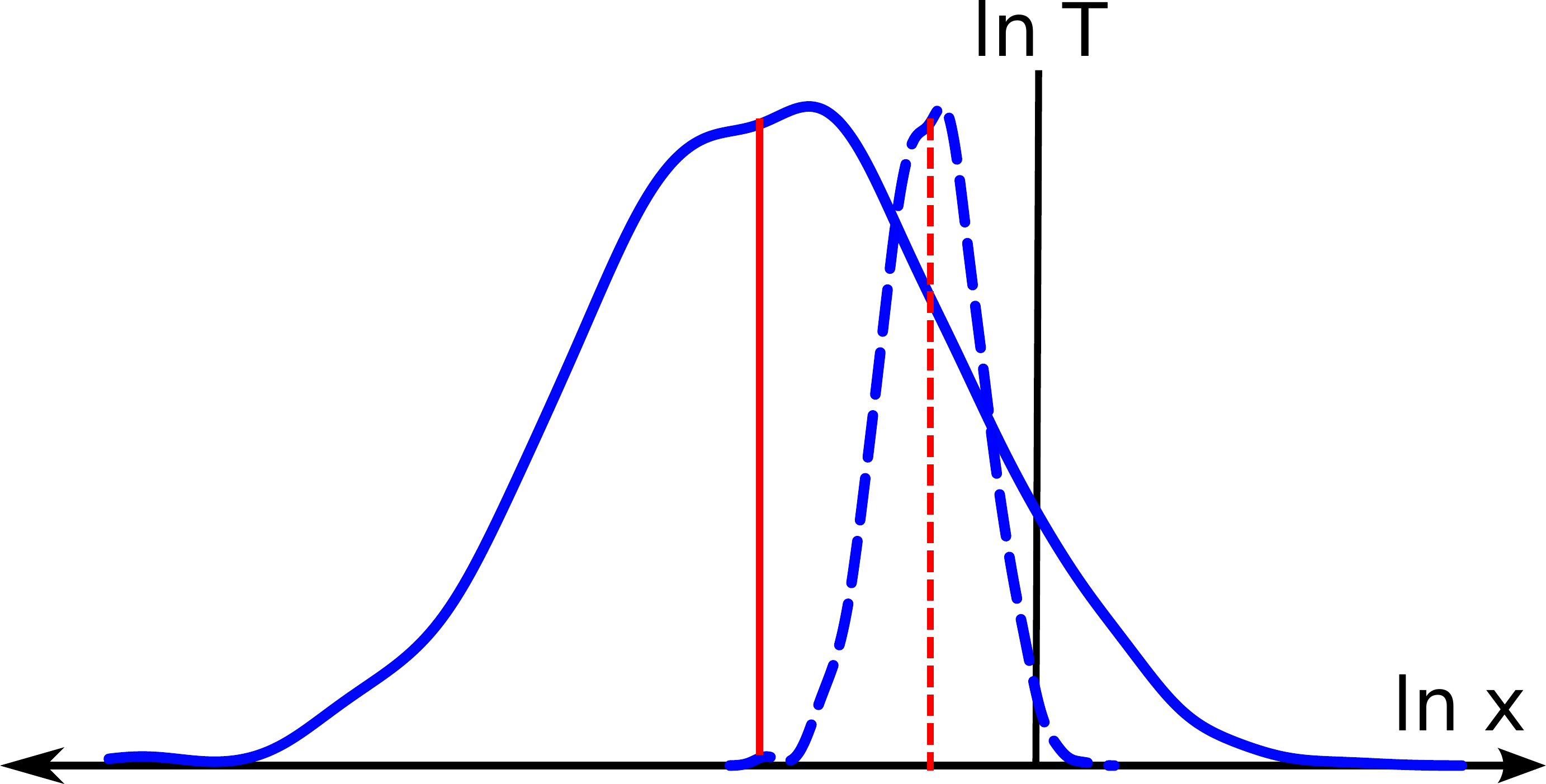}\raisebox{-3ex}{(a)}\hfill 
 \includegraphics[width=0.45\textwidth]{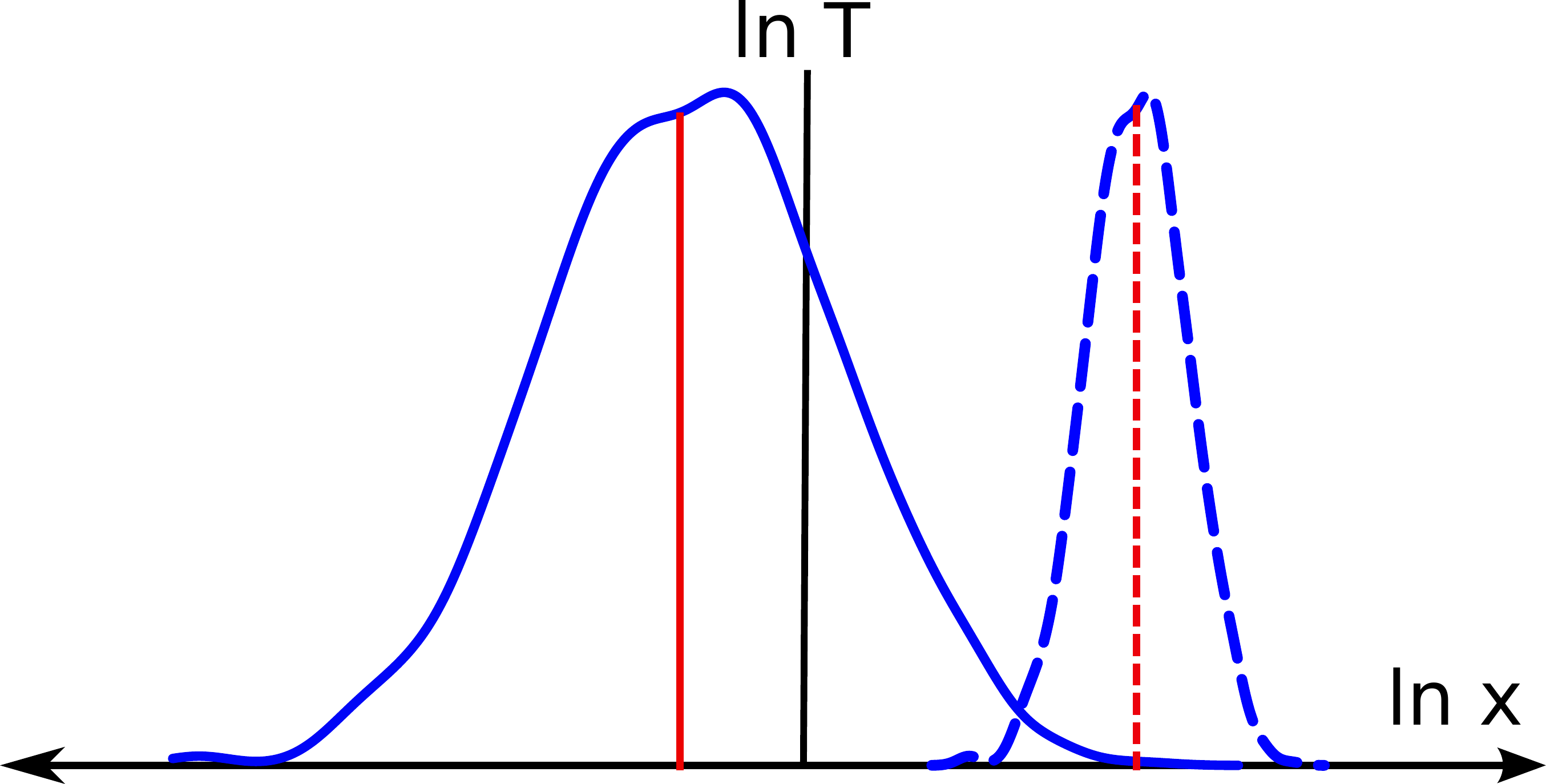}\raisebox{-3ex}{(b)}
 \caption{Sketch of the initial opinion distribution (blue solid line) and the median (red solid line).
   Two different initial conditions (also used in Figure~\ref{woc123-simu}):  \textbf{(a)} $\mathcal{W}(0)$=43, $\mathcal{E}(0)$=0.02, $\ln \mathcal{T}$=-3.12. \textbf{(b)}
    $\mathcal{W}(0)$=46, $ \mathcal{E}(0)$=0.01, $\ln \mathcal{T}=-2.9$. Both: $\mean{\ln x(0)}$=-3.0,
    $\mathcal{D}(0)$=0.72. The position of the final opinion distribution (blue dashed line) and the median (red dashed line) is determined by the parameters $(\alpha,\beta)$. Despite the more favorable initial condition (b), a stronger social influence $\alpha$ leads to a worse outcome.
  }
  \label{woc12}
\end{figure}

\begin{figure}[htbp]
  \centering
  \includegraphics[width=0.45\textwidth]{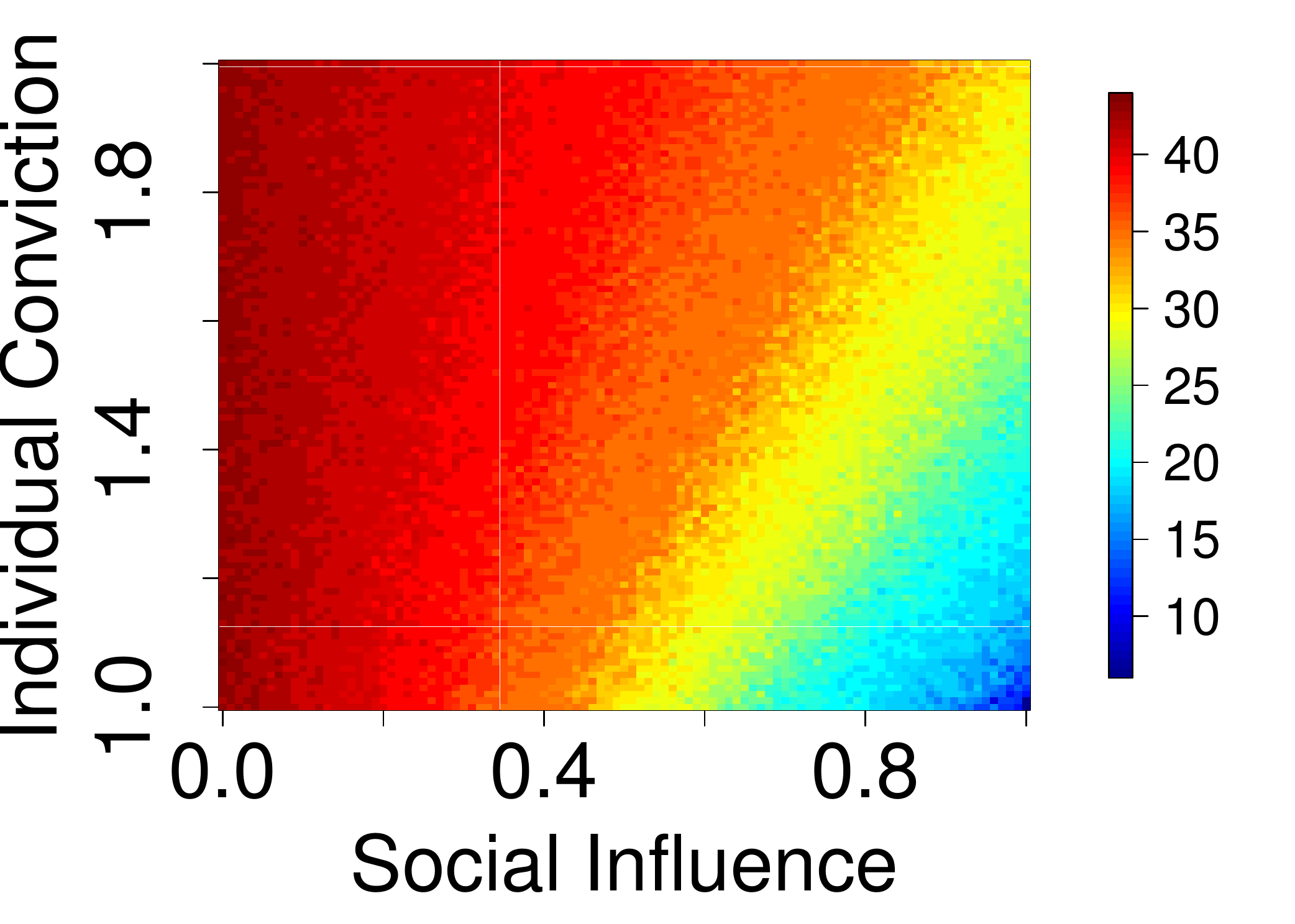}
  \hfill (a) \hfill
\includegraphics[width=0.45\textwidth]{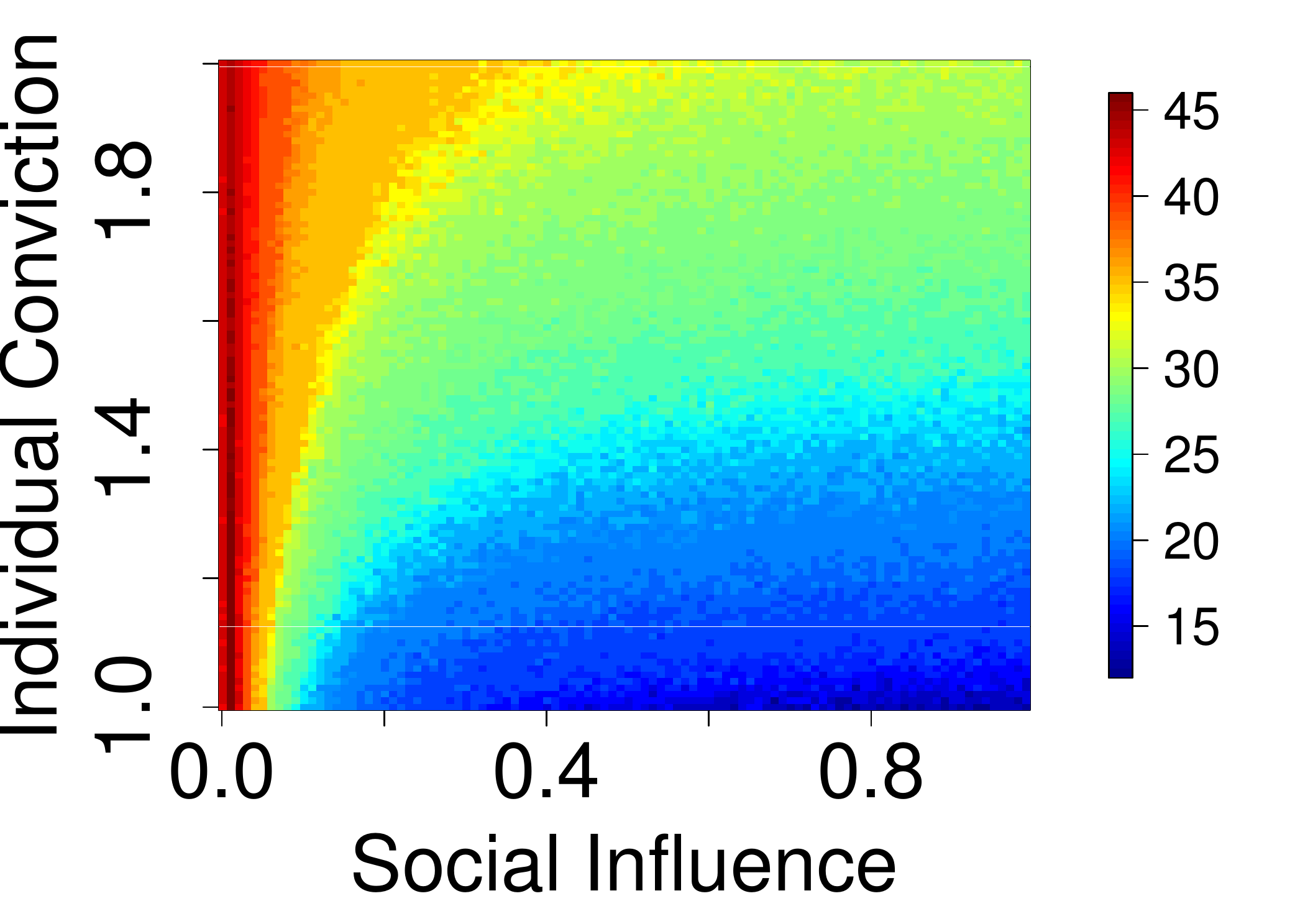}\\
  \includegraphics[width=0.45\textwidth]{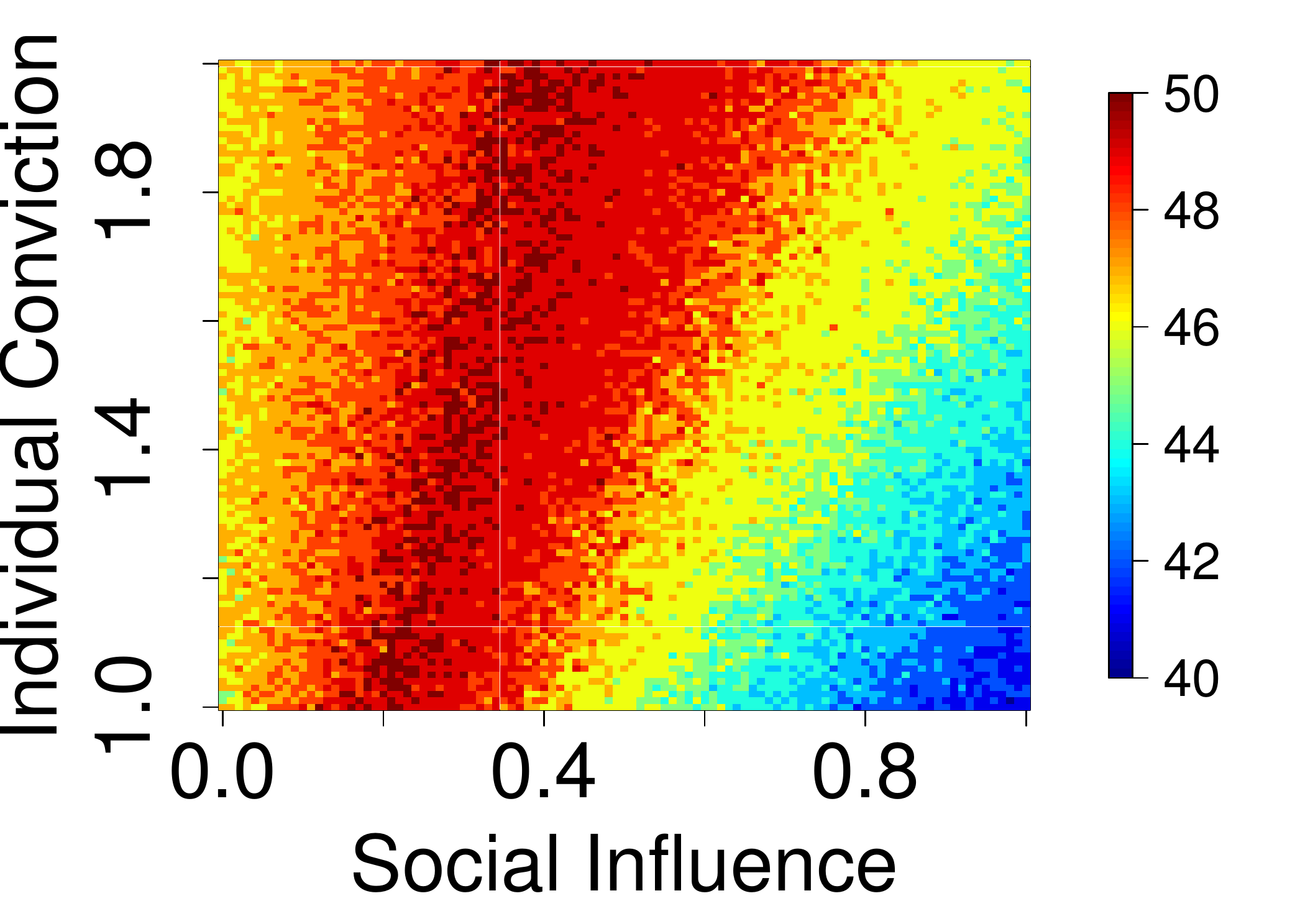}
  \hfill (b) \hfill
  \includegraphics[width=0.45\textwidth]{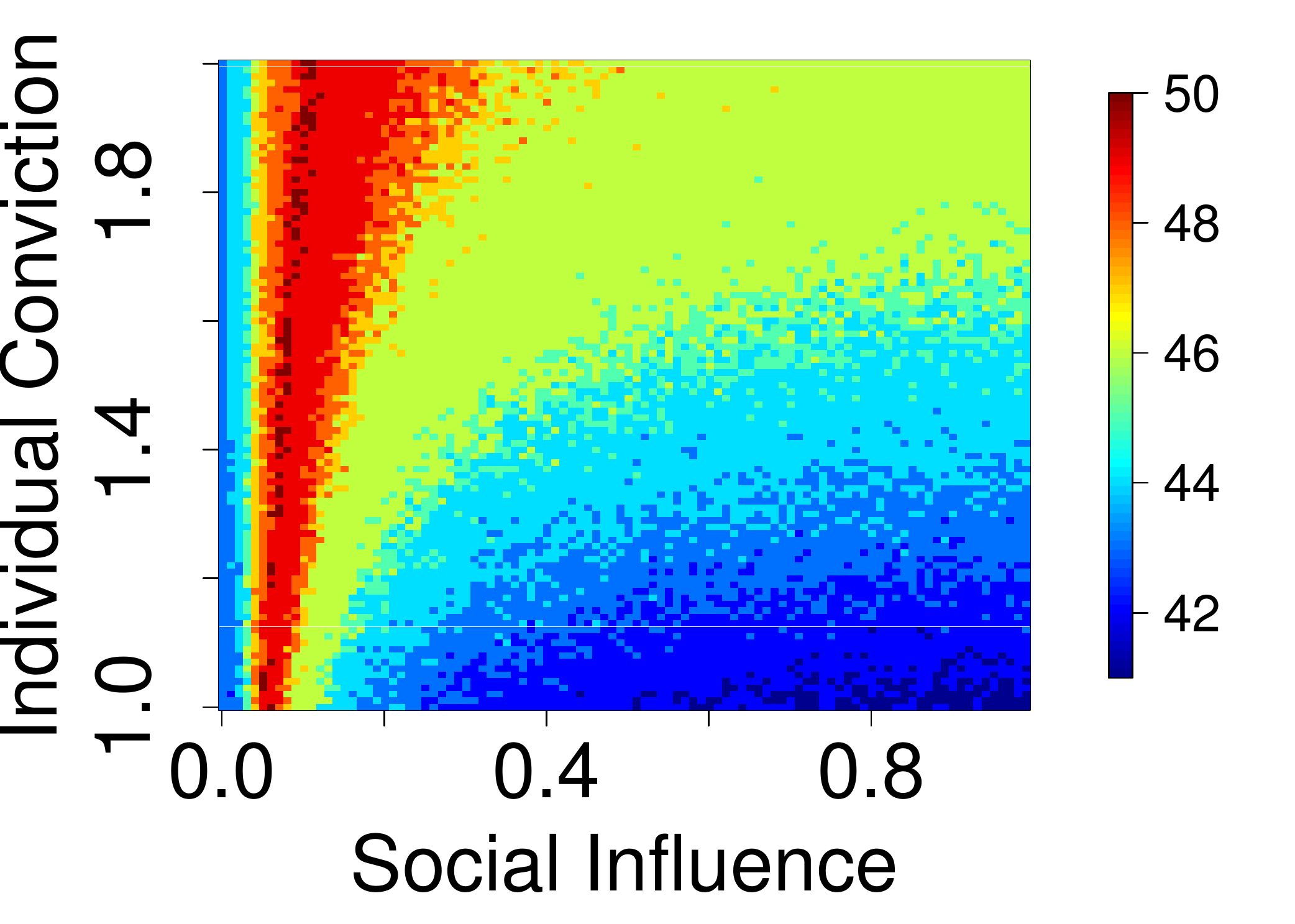}
  \caption{Agent-based simulations of the long-term wisdom-of-crowds indicator $\mathcal{W}_{\mathrm{LT}}$ (color coded) dependent on the values of social influence $\alpha$ ($x$-axis) and individual conviction $\beta$ ($y$-axis).
  \textbf{(left)} Aggregated-information regime, Eqn.~\eqref{estimates}, \textbf{(right)} full-information regime, Eqn.~\eqref{full-info}.
  Parameters: $N$=100, $t$=3000, $A$=$10^{-3}$, $\Delta t$=0.01. 
  Different initial conditions: 
  \textbf{(a)} $\mathcal{W}(0)$=43, $\mathcal{E}(0)$=0.02. \textbf{(b)}
    $\mathcal{W}(0)$=46, $ \mathcal{E}(0)$=0.01. Both: $\mean{\ln x(0)}$=-3.0,
    $\mathcal{D}(0)$=0.72.
  }
  \label{woc123-simu}
\end{figure}

Figure~\ref{woc123-simu} presents the results for the full-information regime and the reference case, for two different initial conditions, which are also sketched in Figure~\ref{woc12}. 
Initial condition (a), which is the same as in Figure~\ref{sweep2}(b), 
has a small initial collective error, but the true value $\ln \mathcal{T}$ is \emph{smaller} than the initial collective opinion $\mean{\ln x(0)}$ (which is not the median, but the mean). 
This implies, in accordance with the discussion of Figure~\ref{sweep2}(b) that the collective opinion cannot converge to the true value. 
Further, $\ln \mathcal{T}$ is not very central with respect to the initial distribution of opinions.
That means the initial WoC indicator is much lower than $[N/2]$=50, which corresponds to the median.
So, in summary, condition (a) reflects rather bad initial condition. 

Initial condition (b), on the other hand, which is the same as in Figure~\ref{sweep2}(c), is comparably better suited for reaching the true value. 
It also has a small initial collective error, but the true value $\ln \mathcal{T}$ is \emph{larger} than the initial collective opinion $\mean{\ln x(0)}$, which means the collective opinion \emph{can} possibly converge to the true value.   
Further, $\ln \mathcal{T}$ is more central with respect to the initial distribution of opinions.
That means the initial WoC indicator is much higher.

Still, as we illustrate in Figures~\ref{woc12},~\ref{woc123-simu}, the more favorable initial condition (b) does not automatically lead to a better wisdom of crowds.
This is due to the opinion \emph{dynamics}, which is rather controlled by the parameters $(\alpha,\beta)$.
Because of the interplay between social influence $\alpha$ and individual conviction $\beta$, agents change their opinions such that (i) the final opinion distribution becomes more narrow, and (ii) the mean $\mean{\ln x_{\mathrm{LT}}}$ shifts toward higher values.
How far it shifts to the right depends on these two social parameters.

As Figure~\ref{woc12} indicates, even the better initial condition can lead to the worse outcome.
This has a systematic and a parameter dependent cause.
As for the systematic one, because the collective opinion shifts toward higher values, also the range of estimates needed to bracket the true value will increase.
This explains why the WoC indicator most often ends up with lower than the initial values.

The parameter dependence displayed in Figure~\ref{woc123-simu} shows that in most cases an increasing social influence $\alpha$ leads to a decreasing WoC indicator (blue region).
This effect again is stronger in the full-information regime.
But an increasing individual conviction is able to mitigate the deterioration.

For the initial condition (b), we observe a non-monotonous dependence of $\mathcal{W}_{\mathrm{LT}}$ on the strength of the social influence, very similar to the dependence shown in Figure~\ref{sweep2}(c) for the collective error.
In this case, rather small values of $\alpha$ enhance the chance to converge to the true value, which is shown by a high WoC indicator.
Higher values of $\beta$ are able to strengthen this effect, enlarging the favorable red region in Figure~\ref{woc123-simu}(b), even in the full-information regime.

So, we can conclude that for both the collective error, Figure~\ref{sweep2}, and the WoC indicator, Figure~\ref{woc123-simu}, we find the same parameter regions where social influence can be beneficial for the wisdom of crowds.
Provided suitable initial conditions, the collective opinion can even converge to the true value, indicated by the minimum collective error, $\mathcal{E}_{\mathrm{LT}}=0$ and the maximum WoC indicator, $\mathcal{W}_{\mathrm{LT}}=[N/2]$.
Despite the loss of group diversity, shown by low values of $\mathcal{D}_{\mathrm{LT}}$ in Figure~\ref{fig:groupdiv}, the crowd becomes ``wiser'' for moderate values of the social influence.
It is non-trivial and interesting to see that all observed effects are much stronger in the full-information regime, in comparison to the reference case.
Because the opinion dynamics is quite sensitive to the available information, the parameter ranges for a favorable outcome considerably decrease in the presence of stronger agent interactions, i.e. in the full-information regime.

\section{Conclusion}
\label{conclusion}

The wisdom of crowds refers to cases in which a larger number of individuals has an opinion about a question for which a true answer exists, although this is not public knowledge.
Then, it was observed already by Galton in 1907 \citep{Galton1907} that the \emph{average opinion} is remarkably close to the true opinion.
This works best if the diversity of individual opinions is large and all opinions are independent.
In most real-world scenarios, however, individuals become aware of the opinions of others and then tend to adjust their own opinion based on this information.
To model such an opinion dynamics is the aim of the current paper.

We have proposed an agent-based model that reflects two competing influences: (i) the individual conviction about the own initial opinion (parameter $\beta$), (ii) the influence of the opinions of others (parameter $\alpha$).
The latter reflects a mutual social influence via the exchange of information. 
Two different regimes are considered: a full-information regime in which agents get to know the opinion of each other agent, and an aggregated-information regime in which agents only get to know the average opinion.
The model assumes that agents have a stronger incentive to adjust their opinion if the difference to other opinions, or to the average opinion, is larger.
This is reasonable: if someone thinks the length of the border between Switzerland and Italy is 500 km and then sees many opinions with values between 2000 and 3000 km, there is a stronger ``force'' to adapt to the majority (even that this is wrong: the true value is 734 km).
This has been confirmed by experiments, already \citep{Lorenz2011a,mavrodiev2013}.
Hence, social influence eventually leads to a decreasing group diversity of opinions.
This is perceived as converging to some sort of consensus, which likely gives more confidence in the outcome of the collective decision process.

Unfortunately, as the same experiments have also shown, because of this convergence crowds easily convince themselves that their collective opinion is the right one, even if it is objectively the wrong one.
Thus, the main objective of our paper is to understand under which conditions this \emph{distorted wisdom of crowds} could emerge.
Our agent-based model plays an valuable role in systematically evaluating these conditions (a) regarding the parameters involved, and (b) regarding the impact of the initial conditions.

Our results demonstrate that the role of social influence cannot be reduced to simplistic messages.
Instead, they reflect that collective opinion dynamics, as most social processes, are more complex.
Specifically, we do find scenarios in which social influence is beneficial for the wisdom of crowds.
If the average initial opinion is far away from the true opinion, i.e. the initial collective error is \emph{large}, social influence can help to reduce the collective error.
But in cases where the initial collective error is already small, social influence mostly distorts the wisdom of crowds, leading to an increase of the collective error.
There is only one situation, where the wisdom of crowds can still be improved, namely if the initial average opinion is below the true opinion and the social influence is very moderate.
The dependency of the WoC effect on the initial conditions is rather critical because, under practical circumstances, these initial
conditions are unknown.
This means, we cannot have a-priori knowledge whether social influence will enhance or distort the WoC.

Our insights help to explain the ``range reduction effect'' found empirically  \cite{Lorenz2011a}.
Because of the adjustment of individual opinions, the opinion distribution, in particular its mean and its median shifts over time such that the true opinion is displaced to peripheral regions of the opinion distribution, while the collective opinion becomes narrowly centered around a wrong value.
This generates a dangerous situation because relying on the average opinion in this case would give the wrong information.
Our investigations show the range of parameters $(\alpha,\beta)$ where social influence generates such outcome, effectively distorting the wisdom of crowds.

Comparing the full-information regime with our reference case, the aggregated-information regime, we note that all effects are much stronger if the information about the opinions of all agents becomes available.
Individual conviction could counterbalance the social influence, but plays a lesser role.
This is also understandable: in the full-information regime, someone not only realizes the differences in opinions, in comparison to the own one, but also how many other individuals deviate from it.
This is hidden in the aggregated-information regime, where someone only knows the difference to the average opinion, hence its impact may be smaller from a psychological perspective.

One could rightly argue against the latter point, because the wisdom of crowds only works with respect to the \emph{average} opinion which should be close to the true opinion. 
So, someone with a basic understanding of the wisdom of crowds should find the average opinion much more reliable, and influential, than knowing all other opinions.
This argument ignores one main point: the wisdom of crowds only works if we have a larger number of \emph{independent} opinions.
Only then, the average opinion may generate the better signal.
Once individuals get to know other opinions, their opinions are no longer adjusted independently, but in response to the social influence generated.
Hence, even a small social influence has the potential to distort the wisdom of crowds, as it was found in experiments and confirmed by our agent-based simulations.

\small \setlength{\bibsep}{1pt}


\begin{thebibliography}{39}
\expandafter\ifx\csname natexlab\endcsname\relax\def\natexlab#1{#1}\fi
\expandafter\ifx\csname url\endcsname\relax
  \def\url#1{\texttt{#1}}\fi
\expandafter\ifx\csname urlprefix\endcsname\relax\def\urlprefix{URL }\fi
\expandafter\ifx\csname selectlanguage\endcsname\relax
  \def\selectlanguage#1{\relax}\fi

\bibitem[{Baldassarri and Bearman(2007)}]{baldassarri2007dynamics}
Baldassarri, D.; Bearman, P. (2007).
\newblock Dynamics of political polarization.
\newblock \emph{American Sociological Review} \textbf{72(5)}, 784--811.

\bibitem[{Banisch(2014)}]{Banisch2014}
Banisch, S. (2014).
\newblock {From Microscopic heterogeneity to macroscopic complexity in the
  contrarian voter model}.
\newblock \emph{Advances in Complex Systems} \textbf{17(5)}, 1450025.

\bibitem[{Banisch and Olbrich(2019)}]{Banisch2019}
Banisch, S.; Olbrich, E. (2019).
\newblock {Opinion polarization by learning from social feedback}.
\newblock \emph{Journal of Mathematical Sociology} \textbf{43(2)}, 76--103.

\bibitem[{Bornschier(2015)}]{bornschier2015new}
Bornschier, S. (2015).
\newblock The New Cultural Conflict, Polarization, and Representation in the
  Swiss Party System, 1975--2011.
\newblock \emph{Swiss Political Science Review} \textbf{21(4)}, 680--701.

\bibitem[{Bose \emph{et~al.}(2017)Bose, Reina and Marshall}]{Bose_2017}
Bose, T.; Reina, A.; Marshall, J.~A. (2017).
\newblock Collective decision-making.
\newblock \emph{Current Opinion in Behavioral Sciences} \textbf{16}, 30–34.

\bibitem[{Castellano \emph{et~al.}(2009)Castellano, Fortunato and
  Loreto}]{castellano2009statistical}
Castellano, C.; Fortunato, S.; Loreto, V. (2009).
\newblock Statistical physics of social dynamics.
\newblock \emph{Reviews of Modern Physics} \textbf{81(2)}, 591.

\bibitem[{Deffuant \emph{et~al.}(2001)Deffuant, Neau, Amblard and
  Weisbuch}]{deffuant2001mixing}
Deffuant, G.; Neau, D.; Amblard, F.; Weisbuch, G. (2001).
\newblock Mixing beliefs among interacting agents.
\newblock \emph{Advances in Complex Systems} \textbf{3}, 87--98.

\bibitem[{Dornic \emph{et~al.}(2001)Dornic, Chat{\'e}, Chave and
  Hinrichsen}]{dornic2001critical}
Dornic, I.; Chat{\'e}, H.; Chave, J.; Hinrichsen, H. (2001).
\newblock Critical coarsening without surface tension: The universality class
  of the voter model.
\newblock \emph{Physical Review Letters} \textbf{87(4)}, 045701.

\bibitem[{Fern{\'a}ndez-Gracia \emph{et~al.}(2014)Fern{\'a}ndez-Gracia,
  Suchecki, Ramasco, San~Miguel and Egu{\'\i}luz}]{fernandez2014voter}
Fern{\'a}ndez-Gracia, J.; Suchecki, K.; Ramasco, J.~J.; San~Miguel, M.;
  Egu{\'\i}luz, V.~M. (2014).
\newblock Is the voter model a model for voters?
\newblock \emph{Physical Review Letters} \textbf{112(15)}, 158701.

\bibitem[{Flache \emph{et~al.}(2017)Flache, M\"{a}s, Feliciani, Chattoe-Brown,
  Deffuant, Huet and Lorenz}]{flache2017}
Flache, A.; M\"{a}s, M.; Feliciani, T.; Chattoe-Brown, E.; Deffuant, G.; Huet,
  S.; Lorenz, J. (2017).
\newblock Models of Social Influence: Towards the Next Frontiers.
\newblock \emph{Journal of Artificial Societies and Social Simulation}
  \textbf{20(4)}, 2.

\bibitem[{Galton(1907)}]{Galton1907}
Galton, F. (1907).
\newblock {Vox Populi}.
\newblock \emph{Nature} \textbf{75(1949)}, 450--451.

\bibitem[{Garcia \emph{et~al.}(2012)Garcia, Mendez, Serdult and
  Schweitzer}]{garcia2012b}
Garcia, D.; Mendez, F.; Serdult, U.; Schweitzer, F. (2012).
\newblock {Political polarization and popularity in online participatory media
  : An integrated approach}.
\newblock In: \emph{Proceedings of the 1st Workshop on Politics, Elections and
  Data - PLEAD '12}. pp. 3--10.

\bibitem[{Groeber \emph{et~al.}(2009)Groeber, Schweitzer and
  Press}]{Groeber2009}
Groeber, P.; Schweitzer, F.; Press, K. (2009).
\newblock {How groups can foster consensus: The case of local cultures}.
\newblock \emph{Journal of Aritifical Societies and Social Simulation}
  \textbf{12(2)}, 4 (1--22).

\bibitem[{Hegselmann and Krause(2002)}]{hegselmann2002opinion}
Hegselmann, R.; Krause, U. (2002).
\newblock Opinion dynamics and bounded confidence models, analysis, and
  simulation.
\newblock \emph{Journal of Artificial Societies and Social Simulation}
  \textbf{5}, 3.

\bibitem[{Holyst \emph{et~al.}(2001)Holyst, Kacperski and
  Schweitzer}]{holyst2001}
Holyst, J.; Kacperski, K.; Schweitzer, F. (2001).
\newblock {Social impact models of opinion dynamics}.
\newblock \emph{Annual Reviews of Computational Physics} \textbf{9}, 253--273.

\bibitem[{Kittur and Kraut(2008)}]{Kittur2008}
Kittur, A.; Kraut, R.~E. (2008).
\newblock Harnessing the wisdom of crowds in wikipedia: quality through
  coordination.
\newblock In: \emph{Proceedings of the 2008 ACM conference on Computer
  supported cooperative work}. CSCW '08, New York, NY, USA: ACM, pp. 37--46.

\bibitem[{Lewenstein \emph{et~al.}(1992)Lewenstein, Nowak and
  Latan\'{e}}]{lewenst-nowak-latane-92}
Lewenstein, M.; Nowak, A.; Latan\'{e}, B. (1992).
\newblock Statistical Mechanics of Social Impact.
\newblock \emph{Physical Review A} \textbf{45}, 703--716.

\bibitem[{Lorenz(2007)}]{lorenz2007}
Lorenz, J. (2007).
\newblock Continuous opinion dynamics under bounded confidence: A survey.
\newblock \emph{International Journal of Modern Physics C} \textbf{18(12)},
  1819-- 1838.

\bibitem[{Lorenz \emph{et~al.}(2011)Lorenz, Rauhut, Schweitzer and
  Helbing}]{Lorenz2011a}
Lorenz, J.; Rauhut, H.; Schweitzer, F.; Helbing, D. (2011).
\newblock {How Social Influence Can Undermine the Wisdom of Crowd Effect}.
\newblock \emph{Proceedings of the National Academy of Sciences of the United
  States of America} \textbf{108(22)}, 9020--9025.

\bibitem[{Mannes(2009)}]{Mannes2009}
Mannes, A.~E. (2009).
\newblock {Are We Wise About the Wisdom of Crowds? The Use of Group Judgments
  in Belief Revision}.
\newblock \emph{Management Science} \textbf{55(8)}, 1267--1279.

\bibitem[{M{\"a}s and Flache(2013)}]{mas2013differentiation}
M{\"a}s, M.; Flache, A. (2013).
\newblock Differentiation without distancing. Explaining bi-polarization of
  opinions without negative influence.
\newblock \emph{PloS ONE} \textbf{8(11)}.

\bibitem[{Mavrodiev and Schweitzer(2020)}]{pm-fs-analytic-20}
Mavrodiev, P.; Schweitzer, F. (2020).
\newblock The ambigous role of social influence on the wisdom of crowds: An
  analytic approach.
\newblock \emph{Physica A} , (under review).

\bibitem[{Mavrodiev \emph{et~al.}(2012)Mavrodiev, Tessone and
  Schweitzer}]{Mavrodiev2012a}
Mavrodiev, P.; Tessone, C.~J.; Schweitzer, F. (2012).
\newblock Effects of social influence on the wisdom of crowds.
\newblock In: \emph{Proceedings of the Conference on Collective Intelligence
  CI-2012}. p. \texttt{https://arxiv.org/html/1204.2991}.

\bibitem[{Mavrodiev \emph{et~al.}(2013)Mavrodiev, Tessone and
  Schweitzer}]{mavrodiev2013}
Mavrodiev, P.; Tessone, C.~J.; Schweitzer, F. (2013).
\newblock {Quantifying the effects of social influence}.
\newblock \emph{Scientific Reports} \textbf{3(1360)}.

\bibitem[{Meng \emph{et~al.}(2018)Meng, Van~Gorder and Porter}]{Meng_2018}
Meng, X.~F.; Van~Gorder, R.~A.; Porter, M.~A. (2018).
\newblock Opinion formation and distribution in a bounded-confidence model on
  various networks.
\newblock \emph{Physical Review E} \textbf{97(2)}, 022312.

\bibitem[{Min and San~Miguel(2017)}]{min2017fragmentation}
Min, B.; San~Miguel, M. (2017).
\newblock Fragmentation transitions in a coevolving nonlinear voter model.
\newblock \emph{Scientific Reports} \textbf{7(1)}, 1--9.

\bibitem[{Nowak \emph{et~al.}(1990)Nowak, Szamrej and
  Latan\'{e}}]{nowak-szam-latane-90}
Nowak, A.; Szamrej, J.; Latan\'{e}, B. (1990).
\newblock From Private Attitude to Public Opinion: A Dynamic Theory of Social
  Impact.
\newblock \emph{Psychological Review} \textbf{97}, 362--376.

\bibitem[{Perony \emph{et~al.}(2013)Perony, Pfitzner, Scholtes, Tessone and
  Schweitzer}]{Pfitzner2013}
Perony, N.; Pfitzner, R.; Scholtes, I.; Tessone, C.~J.; Schweitzer, F. (2013).
\newblock {Enhancing consensus under opinion bias by means of hierarchical
  decision making}.
\newblock \emph{Advances in Complex Systems} \textbf{16(06)}, 1350020.

\bibitem[{Ray(2006)}]{Ray2006}
Ray, R. (2006).
\newblock {Prediction Markets and the Financial "Wisdom of Crowds"}.
\newblock \emph{Journal of Behavioral Finance} \textbf{7(1)}, 2--4.

\bibitem[{Schweighofer \emph{et~al.}(2020{\natexlab{a}})Schweighofer, Garcia
  and Schweitzer}]{schweighofer20}
Schweighofer, S.; Garcia, D.; Schweitzer, F. (2020{\natexlab{a}}).
\newblock An agent-based model of multi-dimensional opinion dynamics and
  opinion alignment.
\newblock \emph{Chaos} , (under review).

\bibitem[{Schweighofer \emph{et~al.}(2020{\natexlab{b}})Schweighofer,
  Schweitzer and Garcia}]{Schweighofer_2020}
Schweighofer, S.; Schweitzer, F.; Garcia, D. (2020{\natexlab{b}}).
\newblock A Weighted Balance Model of Opinion Hyperpolarization.
\newblock \emph{Journal of Artificial Societies and Social Simulation}
  \textbf{23(3)}, 5.

\bibitem[{Schweitzer(2003)}]{agentbook-03}
Schweitzer, F. (2003).
\newblock \emph{{Brownian Agents and Active Particles. Collective Dynamics in
  the Natural and Social Sciences}}.
\newblock Berlin: Springer.

\bibitem[{Schweitzer(2018)}]{schweitzer2018}
Schweitzer, F. (2018).
\newblock {Sociophysics}.
\newblock \emph{Physics Today} \textbf{71(2)}, 40--46.

\bibitem[{Schweitzer(2020)}]{schweitzer2020}
Schweitzer, F. (2020).
\newblock The law of proportionate growth and its siblings: Applications in
  agent-based modeling of socio-economic systems.
\newblock In: H.~Aoyama; Y.~Aruka; H.~Yoshikawa (eds.), \emph{Complexity,
  Heterogeneity, and the Methods of Statistical Physics in Economics}, Tokyo:
  Springer. pp. 145--176.

\bibitem[{Schweitzer and Behera(2009)}]{fs-voter-03}
Schweitzer, F.; Behera, L. (2009).
\newblock {Nonlinear voter models: the transition from invasion to
  coexistence}.
\newblock \emph{The European Physical Journal B} \textbf{67(3)}, 301--318.

\bibitem[{Schweitzer \emph{et~al.}(2020)Schweitzer, Krivachy and
  Garcia}]{Schweitzer_2020}
Schweitzer, F.; Krivachy, T.; Garcia, D. (2020).
\newblock An Agent-Based Model of Opinion Polarization Driven by Emotions.
\newblock \emph{Complexity} \textbf{2020}, 5282035.

\bibitem[{Stark \emph{et~al.}(2008)Stark, Tessone and Schweitzer}]{STARK2008}
Stark, H.~U.; Tessone, C.~J.; Schweitzer, F. (2008).
\newblock {Slower is Faster: Fostering Consensus Formation by Heterogeneous
  Interia}.
\newblock \emph{Advances in Complex Systems} \textbf{11(04)}, 551--563.

\bibitem[{Suchecki \emph{et~al.}(2005)Suchecki, Egu{\'\i}luz and
  San~Miguel}]{suchecki2005vmd}
Suchecki, K.; Egu{\'\i}luz, V.; San~Miguel, M. (2005).
\newblock {Voter model dynamics in complex networks: Role of dimensionality,
  disorder, and degree distribution}.
\newblock \emph{Physical Review E} \textbf{72(3)}, 36132.

\bibitem[{Surowiecki(2005)}]{Surowiecki2005}
Surowiecki, J. (2005).
\newblock \emph{{The Wisdom of Crowds}}.
\newblock Anchor, 336 pp.

\end{thebibliography}
\end{document}